\documentclass[%
 aip,
 amsmath,amssymb,
 reprint,%
]{revtex4-2}

\usepackage{graphicx}
\usepackage{dcolumn}
\usepackage{bm}

\usepackage[utf8]{inputenc}
\usepackage[T1]{fontenc}
\usepackage[%
	breaklinks=true,
	colorlinks=true,
	linkcolor=blue,
	urlcolor=blue,
	citecolor=blue
]{hyperref}
\usepackage{mathptmx}
\usepackage{etoolbox}

\makeatletter
\def\@email#1#2{%
 \endgroup
 \patchcmd{\titleblock@produce}
  {\frontmatter@RRAPformat}
  {\frontmatter@RRAPformat{\produce@RRAP{*#1\href{mailto:#2}{#2}}}\frontmatter@RRAPformat}
  {}{}
}%
\makeatother
\begin{document}

\preprint{AIP/123-QED}

\title[This preprint based on REVTeX 4 template, copyright (c) 2009 American Institute of Physics.]{Simulation aspects of patterning polymer films via evaporative lithography and composite substrates}
\author{Konstantin~S.~Kolegov}
	\email{konstantin.kolegov@asu.edu.ru}
 \affiliation{Mathematical Modeling Laboratory, Astrakhan Tatishchev State University, Astrakhan 414056, Russia}

\date{\today}

\begin{abstract}
The continuing development of evaporative lithography is important for many areas such as the creation of photonic crystals for optronics and microelectronics, the development of biosensors for medical applications and biotechnology, and for the formation of functional coatings for nanotechnology, including the application of thin, protective polymer coatings. The article proposes a mathematical model that allows us to explain the basic mechanisms of the formation of thin polymer films (less than 50~$\mu$m thick) during their deposition onto a composite substrate by  methanol evaporation from a solution. If the thermal conductivity of the substrate is spatially non-uniform, this results in inhomogeneous evaporation along the free film surface. Therefore, as the film dries, a patterned polymer coating is left behind on the substrate. The mathematical model described here is based on the lubrication approximation and takes into account the dependence of the solution density on the concentration. The numerical computation results are in qualitative agreement with the experimental data of other authors. The article shows that solutal Marangoni flow plays a primary role in the process under consideration. This study allows us to better understand the mechanisms that can be used in evaporative lithography.
\end{abstract}

\maketitle

\section{Introduction}
The development of mathematical models, numerical methods, and software is important for various engineering problems~\cite{Xue2023,Xue2024}. The evaporative self-assembly process allows different patterns of deposition to be obtained on a substrate as the droplets and films dry~\cite{Deegan1997,Deegan2000475,Sefiane2014}. A further logical progression of this method is evaporative lithography~\cite{Routh1998,Harris2007,Harris2008}. Various key parameters can be altered to influence the evaporation to produce particular deposit patterns. These approaches can be classified as passive or active methods. In passive methods the parameters are configured ahead of starting process. Active methods are controlled by parameters that can be adjusted in real time. A separate subgroup, of ``hybrid methods'', can be considered,  where evaporative lithography is combined with other methods related, or not, to evaporative self-assembly. A large number of examples are described in the review~\cite{Kolegov2020}. Here we will mention only a few articles in this field, including some relatively recent ones.

When solving problems in the field of evaporative lithography, numerical methods are often used to cope with the nonlinearity of the differential equations, the need to consider several subdomains (liquid, substrate and air) and the complex geometry of the computational domain. However, for some special cases, an analytical solution to the problem may be obtained, for example, the case of an axisymmetric colloidal drop drying on a horizontal impermeable substrate, if we consider the stationary vapor flux density $J$ with a spatial heterogeneity described by an empirical expression using an adjustable parameter~\cite{Ambrosio2023}. This model allows one to calculate the field of the capillary flow affecting the transfer of colloidal particles that depends on the character of the evaporation (for example, $J$ can increase either near the drop periphery or in center, while in other cases evaporation occurs uniformly over the entire free surface). An analytical solution can also be obtained for the Marangoni flow~\cite{Li201972}. For example, such a flow can be controlled by non-uniform heating of the substrate (cell)~\cite{AlMuzaiqer2021126550,AlMuzaiqer2021}. There is also another example: when a temperature gradient arises in a drop of water with suspended gold nanoparticles under local illumination of the free surface. This results in a surface tension gradient and, as a consequence, thermal Marangoni flow develops. Such subregion heating occurs if the light wavelength matches the plasmonic absorption of the gold nanoparticles~\cite{Farzeena2022}. This allows one to control the process of deposit formation in real time and therefore to obtain different patterns,  including of particle mixtures (gold nanoparticles and polymer microspheres) on the substrate. The local irradiation time affects the morphology of such deposits. In addition, the deposit morphology is often affected by the substrate wettability~\cite{PerkinsHoward2022}. A special chemical surface treatment may be used to make a hydrophobic substrate hydrophilic in order to obtain densely packed microspheres instead of disordered structures. This difference is the result of the capillary attraction of particles that occurs on a hydrophilic surface in a thin liquid film. Evaporative self-assembly can also enable microsphere masks to be obtained on a solid surface for further use in colloidal lithography~\cite{PerkinsHoward2022}. In the case of bioliquid droplets, the deposit formation may be controlled using solutal Marangoni flow~\cite{Hegde2022}. When an ethanol drop hanging on the tip of a syringe needle is placed above a sessile drop of bioliquid (above its free surface in a particular, local subregion), a surface tension gradient appears. This results from ethanol vapor being adsorbed at the ``liquid--air'' interface, thereby reducing the bioliquid's surface tension. This method can be used in biomedical applications. The coffee ring effect can be suppressed by directing an infrared laser beam onto the top  of a sessile drop of saline solution, resulting in the formation of a central crystalline deposit~\cite{Li2021a}. The local liquid heating results in a temperature gradient on the free surface and, therefore, in spatially inhomogeneous evaporation. This approach can produce a spot-like deposit formation~\cite{Li2021a}. An experiment~\cite{Goy2022} was performed to measure the resulting spot size versus the power of the IR laser and the diameter of the light beam directed at the top of the sessile water drop. The central spot was surrounded by a ring formed at the contact line. Based on analytical flow velocity estimates, the authors showed that this form of deposit is associated with the competition between the capillary and Marangoni flows~\cite {Goy2022} (stagnation point $r\approx 0.8 R$ where $R$ is the radius of the base of the drop). Another way to achieve the deposition of particles in specific locations is to use a substrate with an array of micropores~\cite{Berneman2021}. If a reduced atmospheric pressure, close to a vacuum, is created under the substrate, the liquid quickly evaporates from the sessile drop through the substrate micropores. A flow of liquid occurs carrying particles towards the pores. To prevent the particles from being sucked into the vacuum region along with the liquid molecules, they must have a size exceeding the pore diameter~\cite{Berneman2021}. Evaporative lithography allows the deposition of functional coatings. For example, electrically conductive ink can be deposited into microchannels of a topographically structured surface through nanoparticle transfer by the capillary flow induced by evaporation~\cite{Corletto2021}. Thermocapillary flow in a liquid film can be affected by spatial inhomogeneity of the substrate's thermal conductivity. The lubrication approximation together with analytical solution have been used to study the influence of a series of parameters on the flow of clear liquid and the shape of the free film's surface on a composite substrate~\cite{GambaryanRoisman2010}. A continuation of that work investigated non-stationary heat distribution in a solid substrate~\cite{GambaryanRoisman2012}. The influence of non-uniform thermal properties and non-uniform substrate heating on thermocapillary flow and on free film surface deformation is described in detail in the review~\cite{GambaryanRoisman2015}. Variable-stiffness composite substrates allow control of the deposit geometry. For example, if such a substrate includes two surface subregions where the outer one is hard and the inner one is soft, then a stripe of deposited particles is formed at the boundary of these subregions~\cite{Iqbal2022}. In this way, stripes may be obtained not only along the perimeters of circles (coffee ring effect) but also along those of squares, triangles, etc. The overall duration of evaporation from a liquid droplet depends on the substrate's thermal conductivity. This parameter can be adjusted in a graphene-polydimethylsiloxane substrate, for example. The higher the graphene concentration in the composite substrate the higher its thermal conductivity, causing the sessile drop to evaporate faster~\cite{Goel2019}. Chemically heterogeneous substrates with uneven distribution of surface wettability (from hydrophilic to hydrophobic) result in instability of liquid nanofilms, up to the appearance of local ruptures. This process can be described using the stochastic lubrication equation and the molecular dynamics method~\cite{Zhao2023}.

Various mechanisms can influence the processes of mass transfer and coating formation: capillary flow~\cite{Deegan1997}, diffusion~\cite{Abe2024}, Marangoni flow~\cite{GambaryanRoisman2015}, etc. The purpose of this article is to study, numerically, the formation of a patterned polymer coating associated with a film drying on a particular composite substrate that had previously been experimentally observed~\cite{Cavadini2013}. Several mechanisms that may presumably influence this process are discussed here. A direct comparison of the numerical results and experimental data~\cite{Cavadini2013} is then used to determine the mechanism playing the primary role.

\section{Methods}

\subsection{Physical problem statement}

\begin{figure}[h]
\includegraphics[width=0.95\columnwidth]{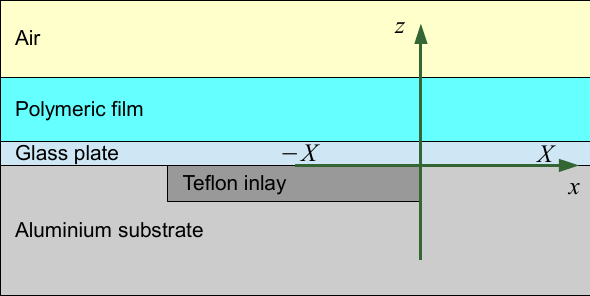}
\caption{\label{fig:sketchCompositeSubstrate}Sketch of the problem statement.}
\end{figure}

Take a liquid polymer film (methanol-poly(vinyl acetate) solution with 67 wt.\% methanol) being applied to a thin glass plate located on top of an aluminum substrate. A thin Teflon inlay is embedded into the aluminum substrate (Fig.~\ref{fig:sketchCompositeSubstrate}). The film height corresponds to the finite size $h$ but its length and width are assumed infinite (the author and readers can imagine this mentally for the convenience of the mathematical description, although in the actual experiment all dimensions were finite). We also assume that the Teflon inlay has a finite width and infinite length. In addition, the height of the Teflon layer is much less than the substrate height so we can assume a zero thickness coating. In consideration of the foregoing, let us consider the problem in a one-dimensional formulation where the film thickness (height) $z=h$ depends on the coordinate $x$ and time $t$. Here, the $y$ axis is disregarded since the film thickness along it will be uniform throughout the process. Heat exchange between the liquid and substrate occurs through the interface between these two phases. It is important to keep in mind that this heat exchange is spatially inhomogeneous, depending on whether the substance in contact with the film is Teflon or aluminum. For this reason, we take into account the substrate thermal conductivity dependence only along the $x$ coordinate.  We assume that the glass plate thickness is much less than that of the aluminum substrate, $h_\mathrm{pl} \ll h_s$. Thus, the thermal properties of the plate do not need to be considered here. The values $h_\mathrm{pl} $ and $h_s$ were not specified in the experiment~\cite{Cavadini2013}. The substrate thickness has been determined approximately based on Fig. 3 of the Ref.~\cite{Cavadini2013} ($h_s\approx$ 2 mm). The initial thickness of the polymer film is known, $h_0\approx$ 150 $\mu$m. The substrate is located perpendicular to the gravity vector. Point $x=-X$ corresponds to the center of the Teflon strip so we consider this to be the boundary of symmetry. The opposite point $x=X$ corresponds to the boundary located at a sufficient distance from the point $x=0$, coinciding with the edge of the Teflon inlay. Let us estimate the time of thermal, $t_T$, and diffusional relaxation, $t_D$, along the vertical direction of the liquid layer, $t_T = h_0^2 / \chi \approx$ 0.2 s and $t_D = h_0^2 / D_0 \approx$ 24.7 s. Note that $t_T  \ll t_D < t_\mathrm{max}$, where $t_\mathrm{max}$ is the evaporation time. Therefore, we disregard the mass and heat transfer along the vertical direction.

The physical and geometric parameters of the problem are presented in Table~\ref{tab:ParametersInCompositeSubstrateProblem}. The values of most of the physical quantities are taken from various handbooks. See Appendix~\ref{secAppendix:parametersForCompositeSubstrateProblem} for more details on the parameters: $\sigma_T$, $c_g$, $\alpha_{sa}$, $\alpha_{ls}$, and $\alpha_{la}$.

\begin{table*}
\caption{Physical and geometric parameters.}
\centering
\begin{tabular}{|p{0.19\linewidth}|p{0.5\linewidth}|p{0.15\linewidth}|p{0.11\linewidth}|}
  \hline
  Symbol&Parameter&Value&UoM\\
  \hline
  $g$& Gravity acceleration & 9.8 & m/s$^2$\\
  $\beta$& Volumetric thermal expansion factor of methanol & $1.26\times 10^{-3}$ & K$^{-1}$\\
  $\beta_\mathrm{air}$& Volumetric thermal expansion factor of air & $3.41\times 10^{-3}$ & K$^{-1}$\\
  $L$& Heat of evaporation & $1162.4\times 10^3$ & J/kg\\
  $C_l$& Specific heat capacity of the liquid & 2619 & J/(kg K)\\
  $C_s$& Specific heat capacity of aluminum substrate & 903 & J/(kg K)\\
  $C_\mathrm{air}$& Specific heat capacity of dry air & 1007 & J/(kg K)\\
  $\sigma_0$& Surface tension coefficient of methanol~\cite{Cavadini2013} & $22\times 10^{-3}$ & N/m\\
  $\eta_0$& Dynamic viscosity of methanol & $0.51 \times 10^{-3}$& s~Pa\\
  $\rho_0$& Methanol density & 792 & kg/m$^3$\\
  $\rho_\mathrm{air}$& Air density & 1.225 & kg/m$^3$\\
  $\nu = \eta_0 / \rho_0$& Kinematic viscosity of methanol & $6.44 \times 10^{-7}$ & m$^2$/s\\
  $\nu_\mathrm{air}$& Kinematic viscosity of air & $15.1 \times 10^{-6}$ & m$^2$/s\\
  $\rho_s$& Aluminum density & $2.7 \times 10^3$ & kg/m$^3$\\
  $k_l$& Thermal conductivity coefficient of the liquid & 0.2 & W/(m K)\\
  $k_a$& Thermal conductivity coefficient of aluminum & 236 & W/(m K)\\
  $k_t$& Thermal conductivity coefficient of Teflon & 0.25 & W/(m K)\\
  $k_\mathrm{air}$& Thermal conductivity coefficient of air & 0.026 & W/(m K)\\
  $\chi= k_l /(C_l \rho_0)$& Thermal diffusion coefficient of the liquid & $9.6 \times
   10^{-8}$ & m$^2$/s\\
  $\chi_\mathrm{air}= k_\mathrm{air} /(C_\mathrm{air} \rho_\mathrm{air})$& Thermal diffusion coefficient of air & $2.1 \times
   10^{-5}$ & m$^2$/s\\
  $c_g$& ``Sol--gel'' phase transition concentration & 0.58 & --\\
  $c_0$& Initial solution concentration~\cite{Cavadini2013} & 0.33 & --\\
  $T_\mathrm{sat}$& Saturation temperature & 338 & K\\
  $T_0$& Ambient temperature~\cite{Cavadini2013} & 303 & K\\
  $\Delta T$& Temperature difference between substrate and the liquid & 2.36 (Table 1~\cite{Ahlers2001}) & K\\
  $\Delta T_{sa}$& Temperature difference between the substrate and air & 10 (Fig. 11~\cite{Cavadini2013}) & K\\
  $\alpha_{ls}$& ``Liquid--substrate'' convective heat exchange coefficient & 61 & W/(m$^2$K)\\
  $\alpha_{la}$& ``Liquid--air'' convective heat exchange coefficient & 50 (see Appendix~\ref{secAppendix:parametersForCompositeSubstrateProblem}) & W/(m$^2$K)\\
  $\alpha_{sa}$& ``Substrate--air'' convective heat exchange coefficient& 0.033 & W/(m$^2$K)\\
  $D_0$& Mutual diffusion coefficient for concentration $c_0$~\cite{Toensmann2021} & $9.1\times 10^{-10}$ & m$^2$/s\\
  $t_\mathrm{max}$& Evaporation process time & 60 & s\\
  $M$ & Molar weight of the liquid & 0.032 & kg/mol\\
  $R$ & Universal gas constant & 8.31 & J/(mol K)\\
  $\rho_p$ & Polymer density & 1190 & kg/m$^3$\\
  $\Delta \rho= \rho_p - \rho_0$ & Density difference between the polymer and liquid & 398 & kg/m$^3$\\
  $\sigma_T = d\sigma/ dT$ & Temperature gradient of surface tension & $-1.782\times 10^{-4}$ & N/(m K)\\
  $2X$& Horizontal region size & 1 & cm\\
  $h_0$& Film thickness (height)~\cite{Cavadini2013} & 150 & $\mu$m\\
  $h_s$& Substrate thickness (height) & 2 & mm\\
  \hline
\end{tabular}
\label{tab:ParametersInCompositeSubstrateProblem}
\end{table*}

As the alcohol evaporates, the film surface cools. A vertical temperature gradient appears in the system. Due to the temperature difference, the substrate loses its heat to the film through thermal conductivity. But, as mentioned earlier, heat transfer will occur non-uniformly along the ``liquid--substrate'' interface, because the substrate is a composite (aluminum conducts heat better than Teflon). The result is that a horizontal temperature gradient appears, promoting non-uniform liquid evaporation along the free film surface. This, in turn, results in the appearance of capillary flows associated with a Laplace pressure gradient. Such a flow is sometimes called compensatory flow since the root cause is non-uniform evaporation, resulting in a curvature change of the free film surface. We neglect the influence of gravity on the fluid flow since a thin liquid film is being considered with a capillary length of $a_c \approx \sqrt{2\sigma_0 / (g \rho_0)}\approx$ 2.4 mm ($h_0 \ll a_c$). Another potential mass transfer mechanism involves thermocapillary flow. Indeed, the temperature gradient along the free surface results in a surface tension difference and, consequently, this leads to the thermal Marangoni effect. The thermocapillary flow is directed along the free surface from the low to the high surface tension regions, and  therefore from the warm region to the relatively cold one. Moreover, the vapor flux density gradient affects the admixture concentration difference along the free liquid surface, also affecting the non-uniform change in surface tension, and, consequently, the appearance of solutal Marangoni flow. Convective and diffusive mass and heat transfer can enhance or weaken the above effects generally, affecting the formation of the solid, patterned polymer coating.

In the experiment~\cite{Cavadini2013}, an air stream was passed over the film with different velocity values, $V_\mathrm{air}=$ 0.5, 1, and 1.5 m/s. The results showed that the profile of the dried polymer film only weakly depended on $V_\mathrm{air}$ in the considered range of values. In this regard, the air flow is not taken into account here.

\subsection{Mathematical model}\label{subsec:ModelInCompositeSubstrateProblem}
The horizontal mass and heat transfer (along the $x$ axis) are important in the system under consideration since diffusion processes smooth out the differences along the vertical direction ($z$ axis), and uniform distribution is expected along the $y$ axis. Let us describe the problem in a one-dimensional formulation. It is convenient to write the mathematical model in dimensionless form to reduce the number of problem parameters (see Table~\ref{tab:DimensionlessParametersInCompositeSubstrateProblem}). The lubrication approximation may be used to describe the hydrodynamics for a thin film ($h_0\approx$ 150~$\mu$m). The liquid is assumed incompressible. In this case, the expression for the horizontal flow velocity averaged over the liquid layer height is written as~\cite{AlMuzaiqer2021}
\begin{equation}\label{eq:averageRadialVelocityForCompositeSubstrateProblem}
	\tilde u = \frac{H(c_g - c)}{\mathrm{Ca}}\left( \frac{\tilde h}{2 \tilde \eta} \frac{\partial \tilde \sigma}{\partial \tilde x} + \frac{\tilde h^2}{3 \tilde \eta} \frac{\partial^3 \tilde h}{\partial \tilde x^3} \right),
\end{equation}
where $\tilde u$, $\tilde h$, $c$, $\tilde \eta$, and $\tilde \sigma$ are the functions depending on the $\tilde x$ coordinate and time $\tilde t$. Here, $c$ is the polymer mass fraction (concentration) in the solution, $\tilde \eta$ is the viscosity, and $\tilde \sigma$ is the surface tension (for information on $\tilde \sigma$ and $\tilde \eta$ see section~\ref{subsec:ClosingRelationsInCompositeSubstrateProblem}). The tilde symbol denotes dimensionless quantities. To convert to dimensional quantities, the following relations should be used: $u = \tilde u u_c$, $h = \tilde h l_c$, $\sigma = \tilde \sigma \sigma_0$, $\eta = \tilde \eta \eta_0$, $t = \tilde t t_c$, and $x = \tilde x l_c$. Here, we use the characteristic length $l_c =$ 1~mm, velocity $u_c =$ 1~mm/s, and time $t_c = $ $l_c / u_c =$ 1~s. In~\eqref {eq:averageRadialVelocityForCompositeSubstrateProblem}, the Heaviside function $H$ is added to eliminate the mass transfer by convective flow in the model when the critical concentration $c_g$ is reached. Formula~\eqref{eq:averageRadialVelocityForCompositeSubstrateProblem} takes into account both capillary flow and Marangoni flow.

\begin{table*}
\caption{Dimensionless parameters.}
\centering
\begin{tabular}{|p{0.25\linewidth}|p{0.59\linewidth}|p{0.15\linewidth}|}
  \hline
  Symbol& Parameter& Value\\
  \hline
  $\mathrm{Ca}=\eta_0 u_c/ \sigma_0$ & Capillary number & $2 \times 10^{-5}$\\
  $\mathrm{Fo}_l=k_l t_c/ (l_c^2 \rho_0 C_l)$ & Fourier number (liquid) & 0.1 \\
  $\mathrm{Fo}_s=k_l t_c/ (l_c^2 \rho_s C_s)$ & Fourier number (substrate) & 0.08 \\
  $\mathrm{Bu}=L/ (C_l T_c)$ & Bulygin number & 1.5 \\
  $\mathrm{Gr}=g l_c^3 \beta \Delta T / \nu^2$ & Grashof number (liquid) & 70 \\
  $\mathrm{Gr}_\mathrm{air}=g l_c^3 \beta_\mathrm{air} \Delta T_{sa} / \nu_\mathrm{air}^2$ & Grashof number (air) & 1.5 \\
  $\mathrm{Pr}=\nu / \chi$ & Prandtl number (liquid) & 6.7 \\
  $\mathrm{Pr}_\mathrm{air}=\nu_\mathrm{air} / \chi_\mathrm{air}$ & Prandtl number (air) & 0.7 \\
  $\mathrm{Ra}= \mathrm{Gr} \mathrm{Pr}$ & Rayleigh number (liquid) & 469 \\
  $\mathrm{Ra}_\mathrm{air} = \mathrm{Gr}_\mathrm{air} \mathrm{Pr}_\mathrm{air}$ & Rayleigh number (air) & 1.05 \\
  $\mathrm{Nu}=C \mathrm{Ra}^n$ & Nusselt number (liquid)\footnotemark[1] & 0.3 \\
  $\mathrm{Nu}_\mathrm{air}=C \mathrm{Ra}_\mathrm{air}^n$ & Nusselt number (air) & $1.26\times 10^{-3}$ \\
  $\mathrm{Gz}_{ls}= C_l \rho_0 l_c / (t_c \alpha_{ls})$& Modified Graetz number (liquid to substrate heat transfer) & 34 \\
  $\mathrm{Gz}_{sl}= C_s \rho_s l_c / (t_c \alpha_{ls})$& Modified Graetz number (substrate to liquid heat transfer) & 40 \\
  $\mathrm{Gz}_{la}= C_l \rho_0 l_c / (t_c \alpha_{la})$& Modified Graetz number (liquid to air heat transfer) & 41.5 \\
  $\mathrm{Gz}_{sa}=C_\mathrm{air} \rho_\mathrm{air} l_c / (t_c \alpha_{sa})$& Modified Graetz number (substrate to air heat transfer) & 37.8 \\
  $\mathrm{Pe}=l_c u_c/ D_0$ & Peclet number & 1099\\
  \hline
\end{tabular}
\footnotetext[1]{Parameters $C$ and $n$ are described in Appendix~\ref{secAppendix:parametersForCompositeSubstrateProblem}.}
\label{tab:DimensionlessParametersInCompositeSubstrateProblem}
\end{table*}

On the other hand, when deriving equations based on the conservation law for the polymer and solution as a whole, we consider the concentration dependence of the liquid density. For more information on the concepts of quasi-incompressible and semi-compressible fluids, see  Ref.~\cite{Roubicek2021}. In this case, the spatiotemporal film thickness evolution is described by the equation
\begin{equation}\label{eq:SolutionConservationLawForCompositeSubstrateProblem}
	\frac{\partial (\tilde h \tilde \rho)}{\partial \tilde t} + \frac{\partial (\tilde h \tilde u \tilde \rho)}{\partial \tilde x} = -\tilde J,
\end{equation}
where the vapor flux density $\tilde J(\tilde x,\tilde t)$ and solution density $\tilde \rho(\tilde x,\tilde t)$ are expressed by the empirical formulas in section~\ref{subsec:ClosingRelationsInCompositeSubstrateProblem}.

The polymer concentration in solution is described by the convection-diffusion equation
\begin{equation}\label{eq:ConvectionDiffusionEqForCompositeSubstrateProblem}
	\frac{\partial c}{\partial \tilde t} + \tilde u \frac{\partial c}{\partial \tilde x} = \frac{H(c_g - c)}{\mathrm{Pe} \tilde \rho \tilde h} \frac{\partial}{\partial \tilde x}\left( \tilde D \tilde h \tilde \rho \frac{\partial c}{\partial \tilde x} \right) + \frac{\tilde J c}{\tilde h \tilde \rho},
\end{equation}
where $\mathrm{Pe}$ is the Peclet number (Table~\ref{tab:DimensionlessParametersInCompositeSubstrateProblem}) and $\tilde D(\tilde x, \tilde t)$ is the mutual diffusion coefficient (section~\ref{subsec:ClosingRelationsInCompositeSubstrateProblem}). In Eq.~\eqref{eq:ConvectionDiffusionEqForCompositeSubstrateProblem}, the Heaviside function $H$ allows us to “turn off” diffusion when a critical concentration $c_g$ is reached. The derivation of the equations~\eqref{eq:SolutionConservationLawForCompositeSubstrateProblem} and \eqref{eq:ConvectionDiffusionEqForCompositeSubstrateProblem} is presented in Appendix~\ref{secAppendix:derivationOfEquationsForCompositeSubstrateProblem}.

Heat transfer in a liquid is described by the equation~\cite{Kolegov2018113}
\begin{multline}\label{eq:HeatTransferEqForCompositeSubstrateProblem}
	\frac{\partial \tilde T_l}{\partial \tilde t} + \tilde u \frac{\partial \tilde T_l}{\partial \tilde x} = \frac{\mathrm{Fo}_l}{\tilde h} \frac{\partial}{\partial \tilde x}\left( \tilde h \frac{\partial \tilde T_l}{\partial \tilde x} \right) -\\ \frac{\tilde J}{\tilde h} (\mathrm{Bu} - \tilde T_l) + \frac{\tilde k_s}{\mathrm{Gz}_{ls}} \frac{\tilde T_s - \tilde T_l}{\tilde h} + \frac{1}{\mathrm{Gz}_{la}} \frac{\tilde T_0 - \tilde T_l}{\tilde h},
\end{multline}
where $\tilde T_l(\tilde x, \tilde t)$ is the liquid temperature ($\tilde T_l = T_l/ T_c$, where $T_c = T_0$ is the characteristic temperature), $\tilde k_s (\tilde x)$ is the thermal conductivity coefficient of the substrate ($\tilde k_s = k_s / k_l$), $\mathrm{Fo}_l$ is the Fourier number for the liquid, $\mathrm{Bu}$ is the Bulygin number, and $\mathrm{Gz}_{ls}$ and $\mathrm{Gz}_{la}$ are the modified Graetz numbers (see Table~\ref{tab:DimensionlessParametersInCompositeSubstrateProblem} and section~\ref{subsec:ClosingRelationsInCompositeSubstrateProblem}). Eq.~\eqref{eq:HeatTransferEqForCompositeSubstrateProblem} takes into account convective and diffusive heat transfers (thermal conductivity), evaporative cooling, and heat exchange via the ``liquid--substrate'' and ``liquid--air'' boundaries. Previously, Eq.~\ref{eq:HeatTransferEqForCompositeSubstrateProblem} was obtained by considering the thermal energy balance in an elementary liquid column under the assumption of constant density~\cite{Kolegov2018113}.

Let us write a separate equation for the substrate temperature~\cite{AlMuzaiqer2021}
\begin{equation}\label{eq:HeatEqForCompositeSubstrateProblem}
	\frac{\partial \tilde T_s}{\partial \tilde t} = \mathrm{Fo}_s \frac{\partial}{\partial \tilde x}\left( \tilde k_s \frac{\partial \tilde T_s}{\partial \tilde x} \right) + \frac{\tilde k_s}{\mathrm{Gz}_{sl}} \frac{\tilde T_l - \tilde T_s}{\tilde h_s} + \frac{1}{\mathrm{Gz}_{sa}} \frac{\tilde T_0 - \tilde T_s}{\tilde h_s},
\end{equation}
where $\tilde T_s(\tilde x, \tilde t)$ is the substrate temperature ($\tilde T_s = T_s/ T_c$), $\mathrm{Fo}_s$~ is the Fourier number for the substrate, and $\mathrm{Gz}_{sl}$ and $\mathrm{Gz}_{sa}$ are the modified Graetz numbers (see Table~\ref{tab:DimensionlessParametersInCompositeSubstrateProblem}). Eq.~\eqref{eq:HeatEqForCompositeSubstrateProblem} takes into account both  the substrate thermal conductivity and the heat exchange with the liquid and the environment. The mathematical model is a system of partial differential equations~\eqref{eq:averageRadialVelocityForCompositeSubstrateProblem}--\eqref{eq:HeatEqForCompositeSubstrateProblem}.

\subsection{Closing relations}\label{subsec:ClosingRelationsInCompositeSubstrateProblem}
The mathematical model described in section~\ref{subsec:ModelInCompositeSubstrateProblem} can be supplemented with closing relations for the functions $\tilde \sigma$, $\tilde \eta$, $\tilde \rho$, $\tilde k_s$, and $\tilde J$. To accomplish this, we will use empirical formulas.

\begin{figure}
	\includegraphics[width=0.9\columnwidth]{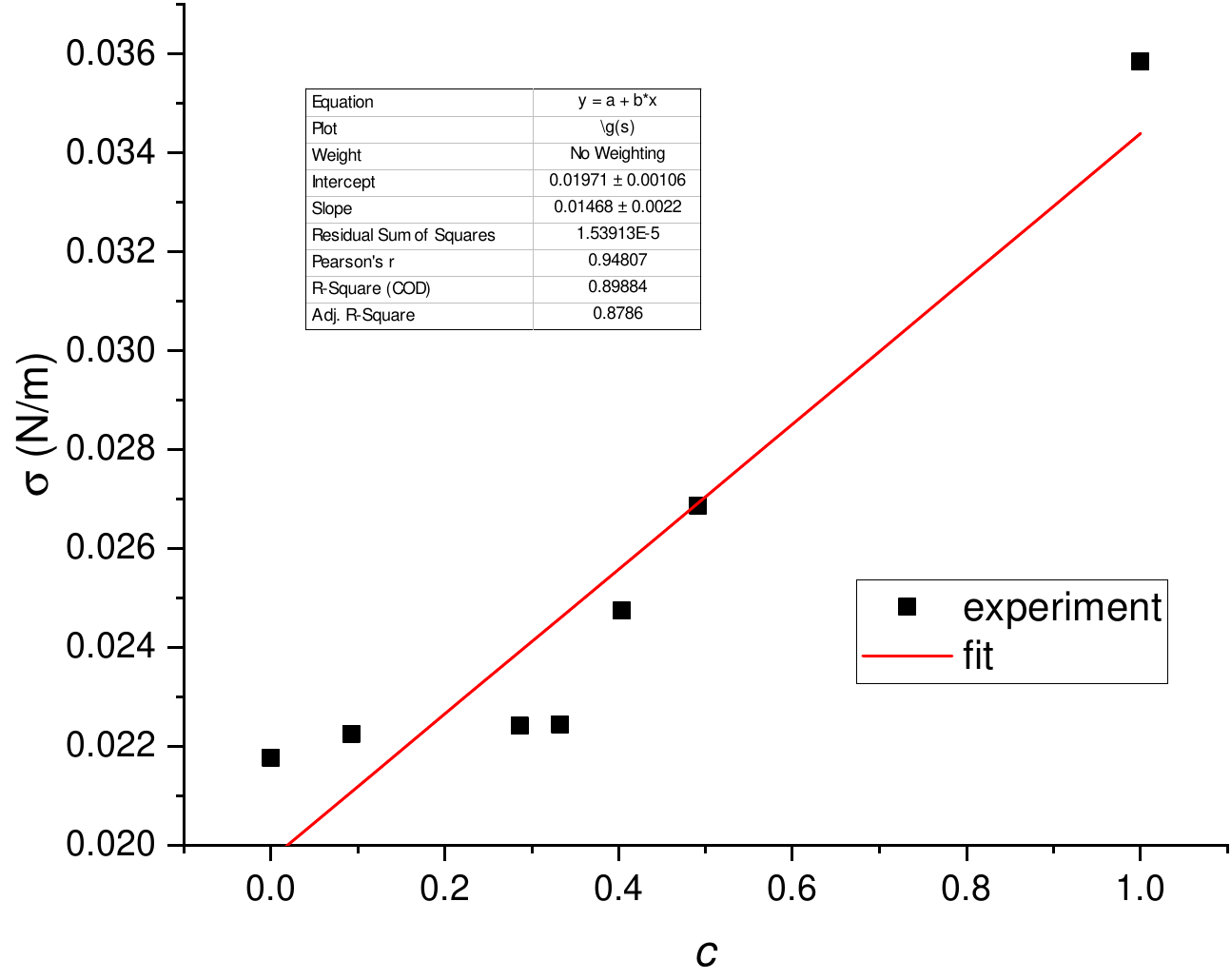}
	\caption{\label{fig:surfaceTensionVsMassFractionInMethanolPolyVinylAcetateSolution}
Dependence of the surface tension on the solution concentration at $T_l \approx$ 303~K.}
\end{figure}

The surface tension depends on the solution temperature and concentration. As a first approximation, we will use a linear dependence
$\tilde \sigma = (a + b c + \sigma_T (T_l - T_0))/\sigma_0$ (Fig.~\ref{fig:surfaceTensionVsMassFractionInMethanolPolyVinylAcetateSolution}). The values of the parameters $a$ and $b$ are chosen to approximate the experimental data~\cite{Cavadini2015} ($a = 0.01971$, $b = 0.01468$).  This dependence allows us to take into account both the solutal Marangoni flow and the thermal one  in formula~\eqref{eq:averageRadialVelocityForCompositeSubstrateProblem}. In order to eliminate the solutal Marangoni flow from the model, it is sufficient to use the value $b=0$. The thermal Marangoni flow may be eliminated from consideration via $\sigma_T = 0$. Capillary flow may be eliminated from the model by multiplying the first term in parentheses in  formula~\eqref{eq:averageRadialVelocityForCompositeSubstrateProblem} by a small number (for example, 10$^{-9}$).

The methanol density varies within 1\% in the considered temperature range, therefore, we take into account only the concentration dependence, $\tilde \rho = (\Delta \rho\, c + \rho_0)/\rho_0$, where $\Delta \rho = \rho_p - \rho_0$. Thus, we also use a linear dependence for density as a first-order approximation.

To describe the dependence of the vapor flux density on liquid temperature, we use the Hertz--Knudsen formula~\cite{Gerasimov2018}
\begin{equation}\label{eq:HertzKnudsenFormulaForPolymerFilm}
	\tilde J=\frac{\alpha_e}{J_c}\sqrt{\frac{M}{2\pi R T_\mathrm{sat}}} (P_\mathrm{sat} - P_v) (1-c^{100}),
\end{equation}
with an additional factor $(1-c^{100})$ added in order to simulate $J\to 0$ at $c\to 1$ in the model (evaporation stops when the liquid runs out). It is difficult to find experimental data on the vapor flux density versus concentration of the solution considered here, but they would be very useful in order to develop a more accurate model. The empirical parameter $\alpha_e$ determines the liquid evaporation rate in  formula~\eqref{eq:HertzKnudsenFormulaForPolymerFilm}. The value $\alpha_e = 7\times 10^{-5}$ is adjusted so that the time of complete liquid evaporation $t_\mathrm{max}$ approximately corresponds to the experimental time~\cite{Cavadini2013}. The characteristic vapor flux density is taken to be $J_c = \rho_0 u_c \approx$ 0.8~kg/(m$^2$s). We assume that the concentration of methanol vapor away from the film is close to zero, therefore, the partial pressure is $P_v\approx 0$~\cite{Sazhin2006}. The saturated vapor pressure $P_\mathrm{sat}$ depends on temperature. This dependence can be described using the semi-empirical Antoine equation, $$P_\mathrm{sat} \approx 133.3\times 10^{A_p - \frac{B_p}{C_p + T_l - 273.15}},$$ where the constants 133.3 and 273.15 are used to convert from one unit of measurement to another (from mm Hg to Pa and from $^\circ$C to K). The following empirical parameter values are used for methanol: $A_p =$ 8.08097, $B_p =$ 1582.27, and $C_p = $ 239.7.

\begin{figure} \includegraphics[width=0.85\columnwidth]{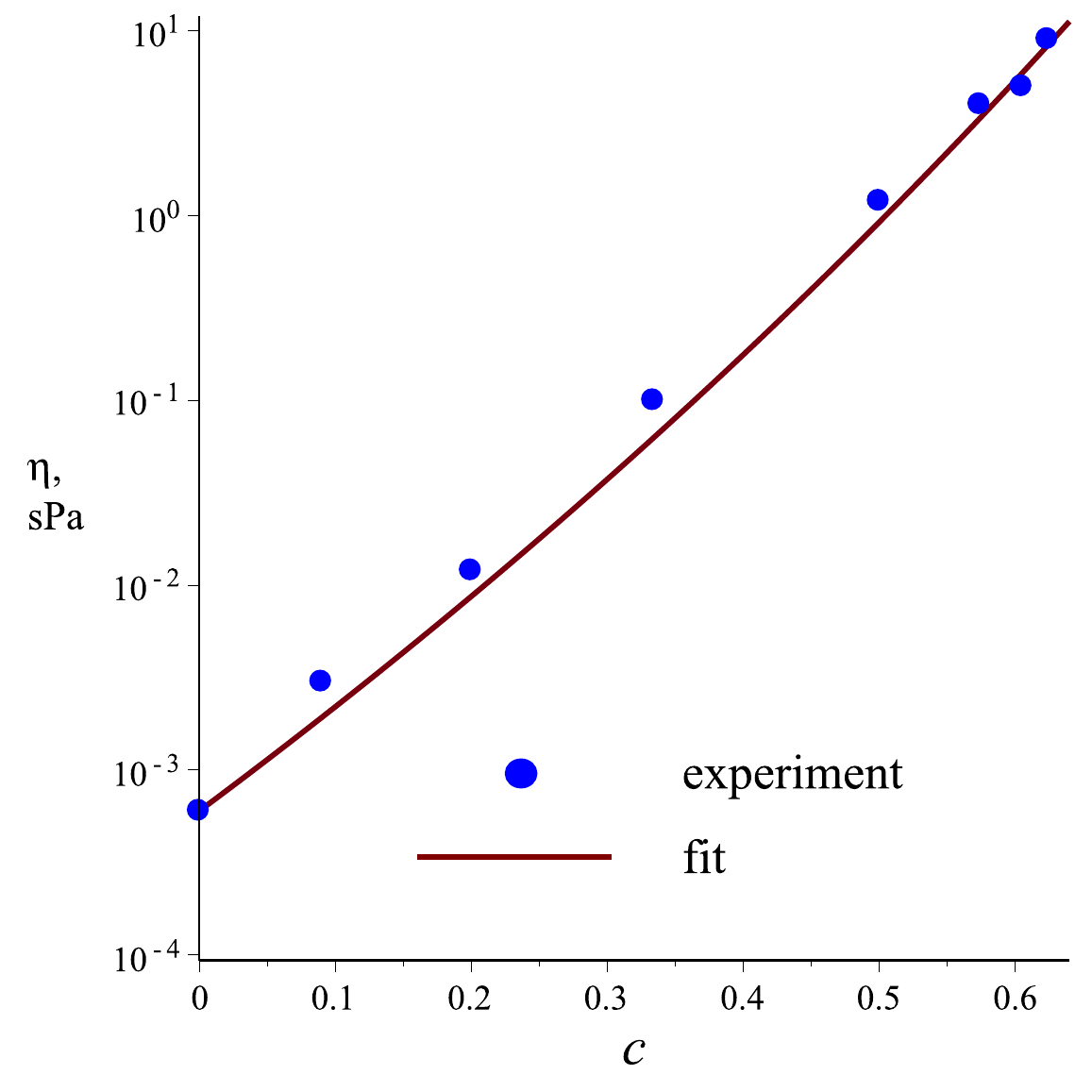}
	\caption{\label{fig:viscosityMethanolPolyVinylAcetateSolution}Dependence of the solution viscosity on the concentration at $T_l =$ 30~$^\circ$C (logarithmic scale on the vertical axis; the circles denote experimental data~\cite{Toensmann2021}, the line relates to the Mooney formula).}
\end{figure}

The methanol viscosity varies within 6\% in the considered temperature range, which is very insignificant. Therefore, we will only take into account the dependence of viscosity on solution concentration. To accomplish this, we use the Mooney formula~\cite{Tarasevich2011} 
\begin{equation}\label{eq:MooneyFormulaForPolymerFilm}
	\eta = \eta_0 \exp \left( \frac{S c}{1 - K c} \right),
\end{equation}
where $S=12.62$ and $K=0.283$. The empirical parameters $S$ and $K$ are determined using the least squares method (Fig.~\ref {fig:viscosityMethanolPolyVinylAcetateSolution}). The exponent will return huge values at high concentrations so we can multiply its argument by the Heaviside function, $$\frac{S c H(c_g - c)}{1 - K c},$$ to minimize computational problems in~\eqref{eq:MooneyFormulaForPolymerFilm}. The fact is that, at $c > c_g$, the viscosity $\eta \to \infty$. Due to the Heaviside function, the expression for the flow velocity~\eqref{eq:averageRadialVelocityForCompositeSubstrateProblem} gives the value 0 at $c > c_g$ so the viscosity calculation result is no longer so important.

The mutual diffusion coefficient depends on the solution concentration, 
\begin{equation}\label{eq:DiffusionForPolymerFilm}
	\tilde D= \frac{D}{D_0} = \exp\left( - \frac{A_d + B_d (1-c)/c}{1 + C_d (1-c)/c} \right).
\end{equation}
Let us take the following values of empirical parameters in~\eqref{eq:DiffusionForPolymerFilm}: 	$A_d=$ 30.39,
$B_d=$ 111.17, and $C_d=$  5.57 (see Table A4~\cite{Toensmann2021}).

The thermal conductivity coefficient of the substrate is a function of a spatial coordinate, $k_s (x) \approx k_t + H_\mathrm{an}(x) k_a$. This approximation is written considering that $k_t \ll k_a$. Here, to smooth out the sharp drop in thermal conductivity at point $x = 0$, the analytical Heaviside function $$H_\mathrm{an}(x) = \frac{1}{1+\exp(-2\kappa x)}$$ is used, where $\kappa=10 / \Delta x$, and the parameter $\Delta x$ controls the transition zone size (the value $\Delta x = 0.3$ was used in the calculations).

\subsection{Initial and boundary conditions}
Assume the shape of the free surface is flat at the initial time, $\tilde h(\tilde x, \tilde t=0)= \tilde h_0$, where $\tilde h_0 = h_0 / l_c$, and, in addition, that temperature and concentration are distributed uniformly, $\tilde T_l(\tilde x, \tilde t=0)= \tilde T_s(\tilde x, \tilde t=0) = \tilde T_0 = 1$ and $c(\tilde x, \tilde t=0)= c_0$. At the boundary $\tilde x = -\tilde X$, let us use the following boundary conditions for symmetry reasons: $$\frac{\partial \tilde h}{\partial \tilde x}= \frac{\partial c}{\partial \tilde x}= \frac{\partial \tilde T_l}{\partial \tilde x}= \frac{\partial \tilde T_s}{\partial \tilde x}= \tilde u= 0.$$ In the zero approximation, suppose that the flow velocity, film height, concentration, and temperature reach constant values away from the difference in substrate thermal conductivity ($\tilde x = 0$) at the boundary $\tilde x = \tilde X$, so let's write the boundary conditions as $$\frac{\partial \tilde h}{\partial \tilde x}= \frac{\partial \tilde u}{\partial \tilde x}= \frac{\partial c}{\partial \tilde x}= 0, \, \tilde T_s = \tilde T_l = \tilde T_0.$$

\subsection{Numerical methods}
 The problem was solved by the finite difference method. The pdsolve module of the Maple 18 was used for the numerical computation. Central differences were used to discretize the spatial derivatives. The implicit difference scheme with first and second order approximations in time and space, respectively, was solved by Newton's iterative method. The computation used a spatial grid consisting of $N+1$ nodes ($N=200$). In this case, the spatial step is $\tilde \xi = 2\tilde X/N= 0.05$. To perform the computational experiments, the time step $\tilde \tau = 0.5$ was used. The model was verified by checking the implementation of the mass conservation law for the polymer. To accomplish this, the expression $$\sum_{i=0}^N h_i^j c_i^j \rho_i^j$$ was computed at the initial and final time steps. Here, $i$ ($j$) is a number representing the spatial (time) node. The error was about 1\%, which may be due to the computational accuracy. To validate the model the numerical results and experimental data were compared (see section~\ref{sec:ResultsAndDiscussionCompositeSubstrateProblem}).

 \section{Results and discussion}\label{sec:ResultsAndDiscussionCompositeSubstrateProblem}
\begin{figure*}
 	\begin{minipage}[h]{0.49\linewidth}
 		\center{\includegraphics[width=0.6\linewidth]{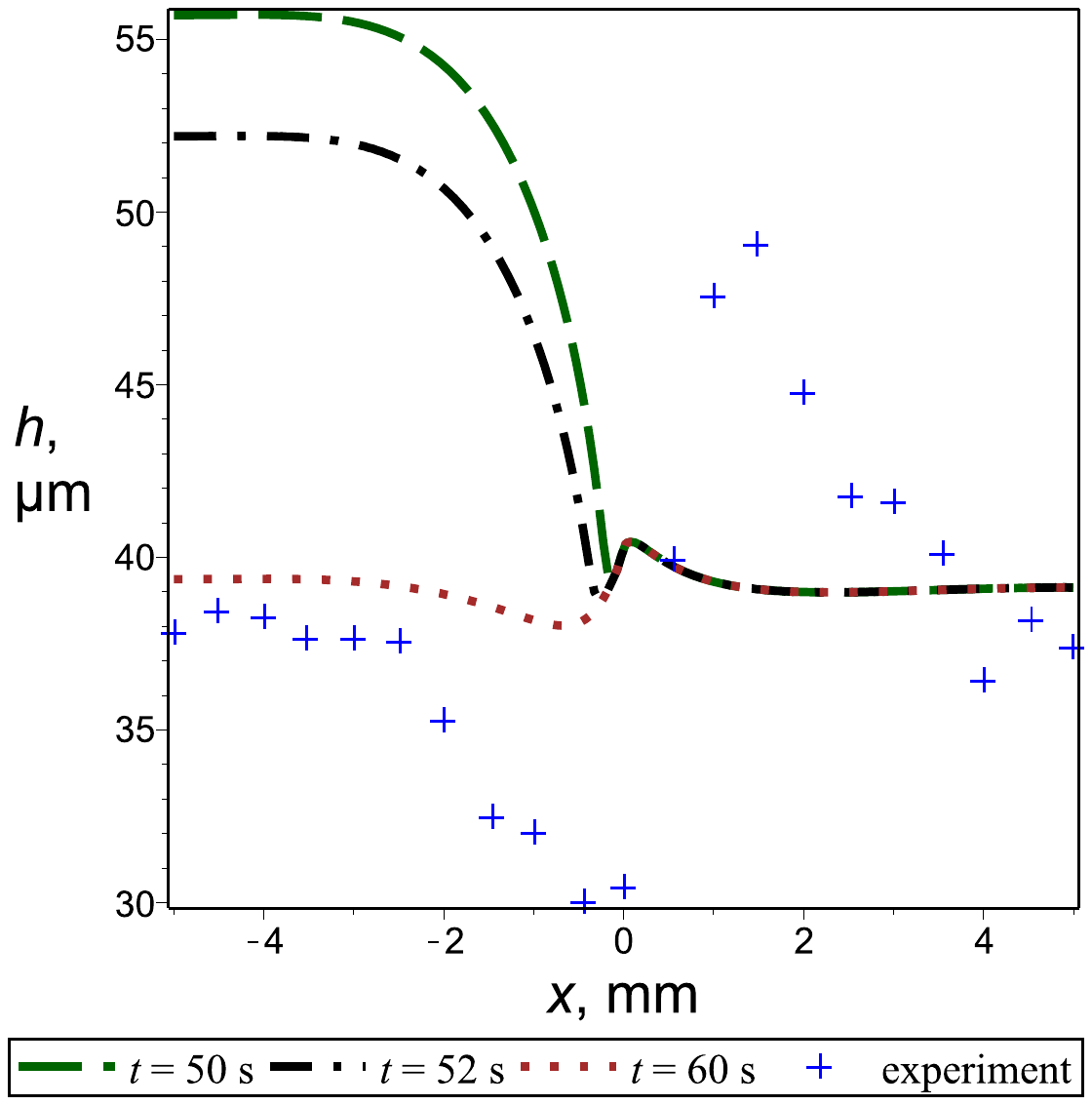} \\ (a)}
 	\end{minipage}
 	\hfill
 	\begin{minipage}[h]{0.49\linewidth}
 		\center{\includegraphics[width=0.6\linewidth]{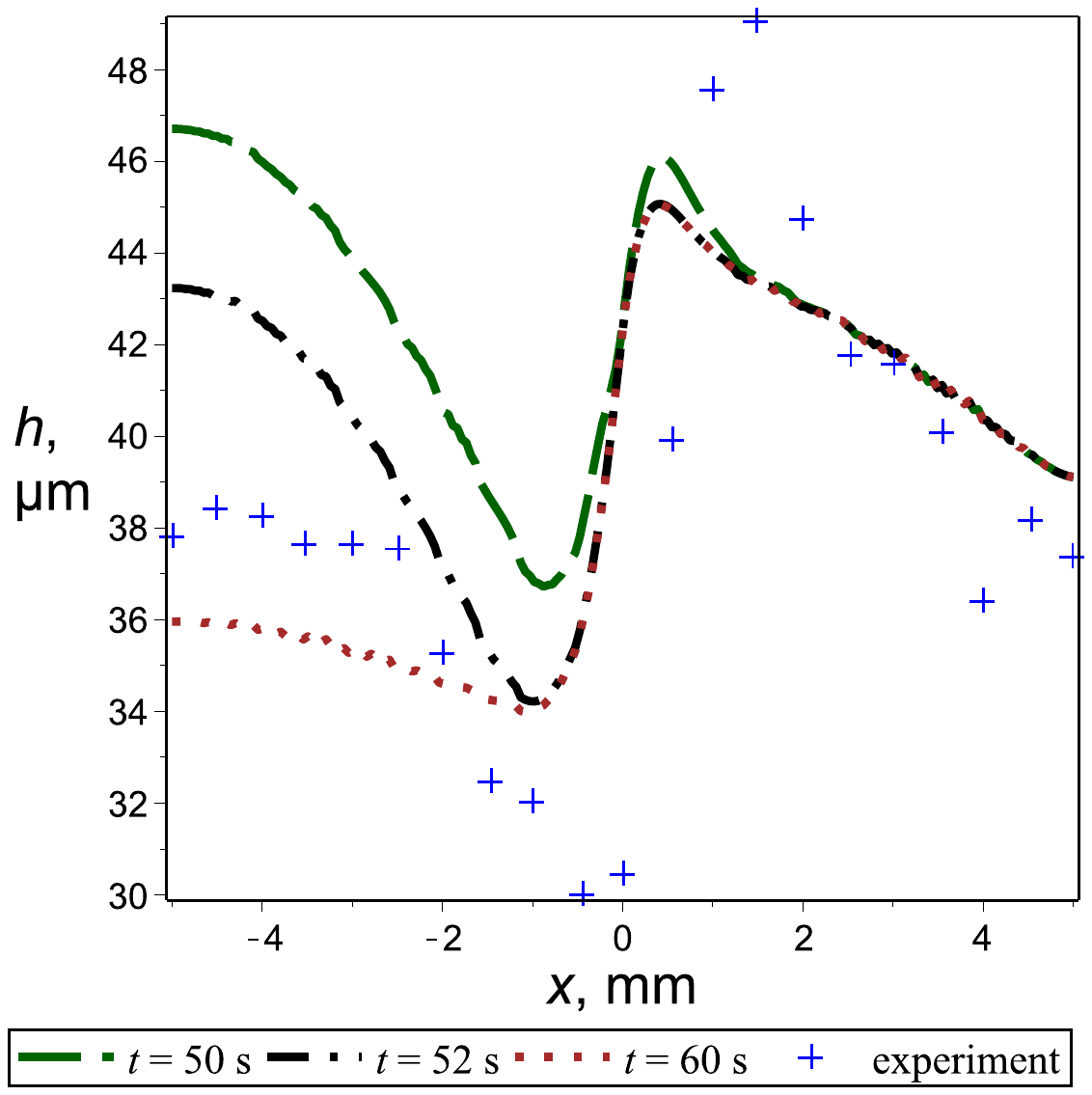} \\ (b)}
 	\end{minipage}
 	\caption{\label{fig:ReliefPolymerFilmForCompositeSubstrateProblem} The polymer coating shape at the end of methanol evaporation (simulation vs. experiment). The model takes into account: (a) the capillary flow and the thermal Marangoni flow; (b) the capillary flow and the solutal Marangoni flow.}
 \end{figure*}

 To study the formation of a patterned polymer coating during alcohol evaporation, a series of computational experiments were performed. From these, first of all, it is necessary to understand which mechanism plays the primary role: the thermal Marangoni flow, the capillary flow or the solutal Marangoni flow. To accomplish this, let us compare the results of two computations with the experimental data~\cite{Cavadini2013}. The first computation ``A'' takes into account the capillary flow and the thermal Marangoni flow (Fig.~\ref{fig:ReliefPolymerFilmForCompositeSubstrateProblem}a).  The solutal Marangoni  flow and the capillary flow are taken into account in the second computation ``B'' (Fig.~\ref{fig:ReliefPolymerFilmForCompositeSubstrateProblem}b). If capillary flow dominated, then both results would be similar but this is far from the case. We do not know what the residual fraction of liquid was when measuring the final film thickness~\cite{Cavadini2013}, therefore, Fig.~\ref{fig:ReliefPolymerFilmForCompositeSubstrateProblem} shows numerical results for several time points towards the end of the evaporation process. The comparison presented in Fig.~\ref{fig:ReliefPolymerFilmForCompositeSubstrateProblem}b shows the qualitative similarity of the numerical computation results and the experimental data (see Fig.~9 in Ref.~\cite{Cavadini2013}). This allows us to conclude that solutal Marangoni flow in the system under consideration plays a primary role. For this reason, the results presented below relate solely to computation ``B''. Perhaps the quantitative match did not work out due to assuming $V_\mathrm{air}= 0$  when constructing the model, but the experimental data~\cite{Cavadini2013} correspond to the value $V_\mathrm{air}= 0.5$ m/s in Fig.~\ref{fig:ReliefPolymerFilmForCompositeSubstrateProblem}. The accuracy of numerical calculations has been determined using error control in Maple 18. The error does not exceed 3\% for the final shape of the film surface (Fig.~\ref{fig:ReliefPolymerFilmForCompositeSubstrateProblem}b).

\begin{figure*}
	\begin{minipage}[h]{0.49\linewidth}
		\center{\includegraphics[width=0.65\textwidth]{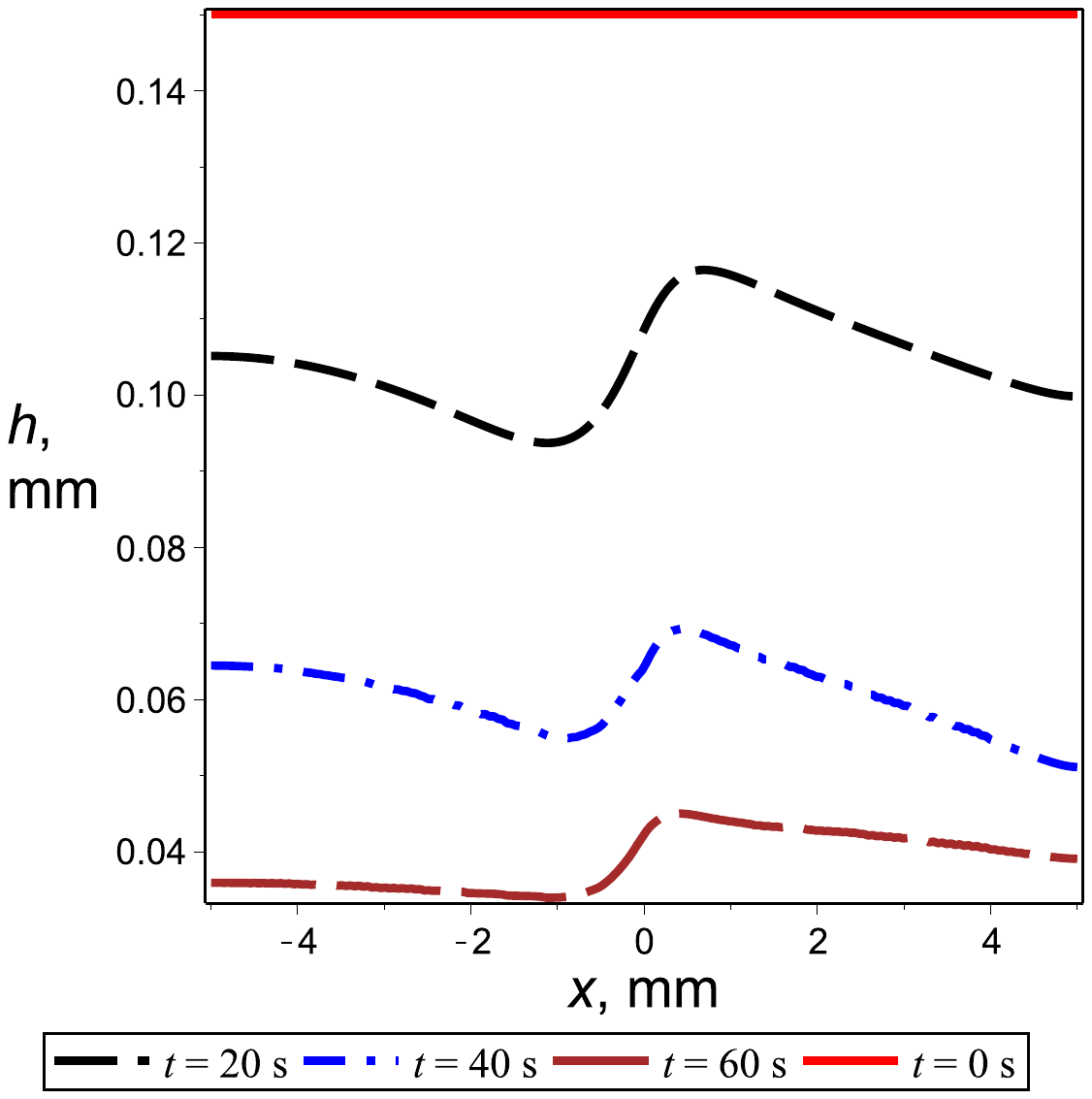} \\ (a)}
	\end{minipage}
	\hfill
	\begin{minipage}[h]{0.49\linewidth}
		\center{\includegraphics[width=0.7\textwidth]{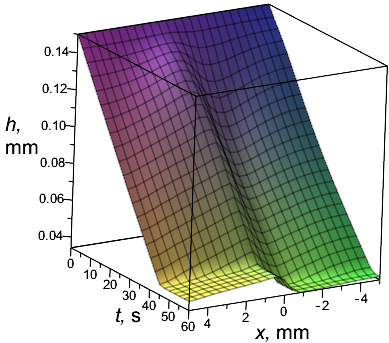} \\ (b)}
	\end{minipage}
	\caption{\label{fig:FilmShapeForCompositeSubstrateProblem} Shape of the free  surface of the film: (a) 2D plots for several consecutive time points; (b) spatiotemporal evolution of $h$ (3D visualization).}
\end{figure*}

The shape of the free film surface $h$ for several consecutive time points is shown in Fig.~\ref{fig:FilmShapeForCompositeSubstrateProblem}a. In addition, for convenience, the spatiotemporal evolution of $h$ is shown in the form of a 3D plot (Fig.~\ref{fig:FilmShapeForCompositeSubstrateProblem}b). The shape of the free film surface at the initial time is flat. Further, during the evaporation, this surface shape becomes curved. The film height decreases over time as the liquid evaporates. As in the experiment~\cite{Cavadini2013}, reduction of $h$ in the Teflon region ($x<0$) can be observed, and thickening of this layer occurs in the aluminum region ($x>0$). The fact is that the polymer weight increases over time in the aluminum region and decreases over the Teflon since the fluid flow transfers the admixture from the Teflon to the aluminum.
\begin{figure} \includegraphics[width=0.65\columnwidth]{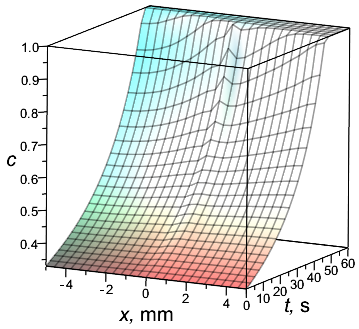} \caption{\label{fig:MassFractionEvolutionForCompositeSubstrateProblem}Spatiotemporal evolution of the polymer mass fraction.}
\end{figure}

The polymer mass fraction increases with time above the aluminum faster than above the Teflon (Fig.~\ref{fig:MassFractionEvolutionForCompositeSubstrateProblem}). After the 50$^\mathrm{th}$ second, not much fluid remains, and that which does is principally above the Teflon, that is, evaporation no longer occurs above the  aluminum. By the 60$^\mathrm{th}$ second of the process, evaporation stops completely since no alcohol remains in the system. At this moment, a solid patterned polymer film has formed. Around point $x=0$ a small bump may be noticed in the mass fraction value (Fig.~\ref {fig:MassFractionEvolutionForCompositeSubstrateProblem}), it is there that the valley is adjacent to the peak of $h$ (Fig.~\ref{fig:ReliefPolymerFilmForCompositeSubstrateProblem}b).
\begin{figure*}
	\begin{minipage}[h]{0.49\linewidth}
		\center{\includegraphics[width=0.65\columnwidth]{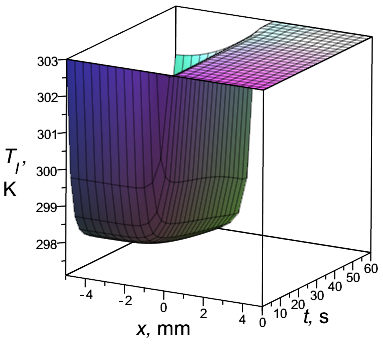} \\ (a)}
	\end{minipage}
	\hfill
	\begin{minipage}[h]{0.49\linewidth}
		\center{\includegraphics[width=0.65\columnwidth]{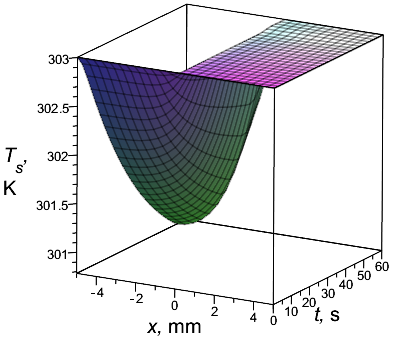} \\ (b)}
	\end{minipage}
	\caption{\label{fig:TemperatureEvolutionForCompositeSubstrateProblem} Spatiotemporal temperature evolution: (a) liquid; (b) substrate.}
\end{figure*}
\begin{figure*}
	\begin{minipage}[h]{0.49\linewidth}
		\center{\includegraphics[width=0.65\textwidth]{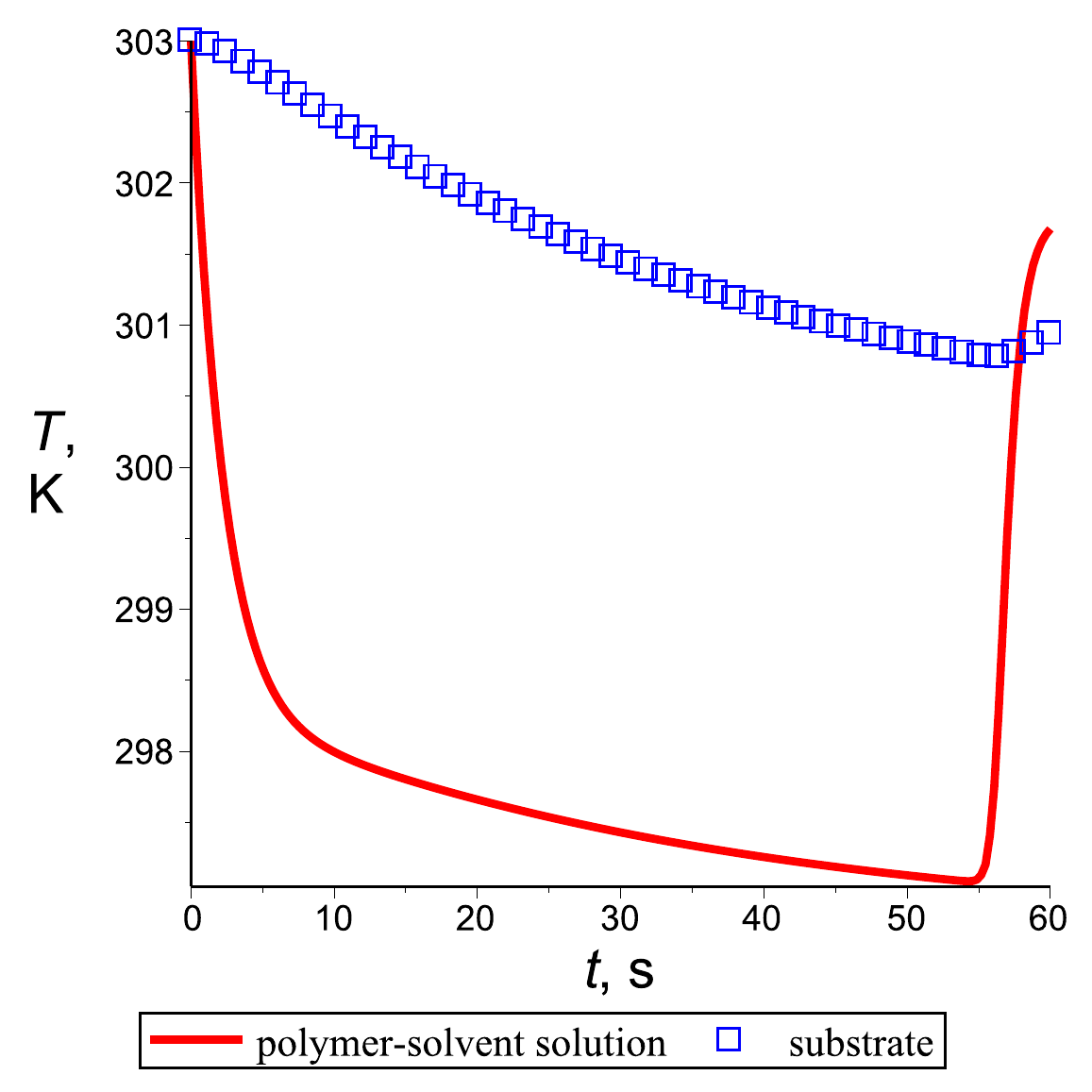} \\ (a)}
	\end{minipage}
	\hfill
	\begin{minipage}[h]{0.49\linewidth}
		\center{\includegraphics[width=0.65\textwidth]{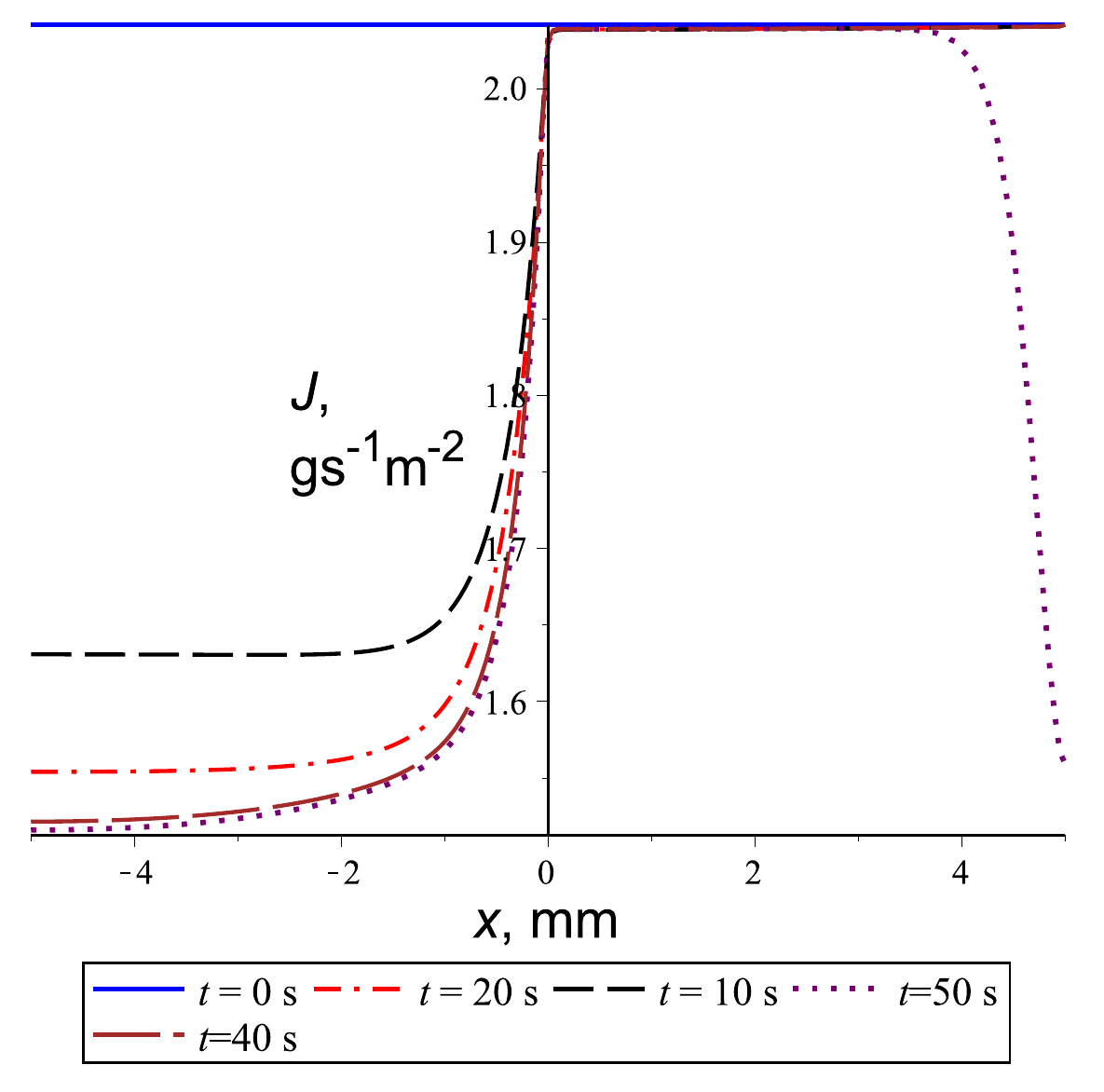} \\ (b)}
	\end{minipage}
	\caption{\label{fig:TemperatureComparisonAndEvaporationRateForCompositeSubstrateProblem} (a) Comparison of the substrate and liquid temperatures over time for point $x=-X$. (b) Spatial dependence of vapor flux density for several consecutive time points.}
\end{figure*}

The temperature calculation results indicate that the film (Fig.~\ref{fig:TemperatureEvolutionForCompositeSubstrateProblem}a) and substrate temperature (Fig.~\ref{fig:TemperatureEvolutionForCompositeSubstrateProblem}b) in the aluminum region remain almost unchanged over time. Aluminum has a relatively high thermal conductivity coefficient so the heat transferred by this material to the solution is quickly balanced by the environmental heat. The temperature in the region of the Teflon, which has a relatively low thermal conductivity, drops over time since evaporative cooling does not have time to be balanced by heat inflow from outside.
\begin{figure}[h]
	\includegraphics[width=0.7\columnwidth]{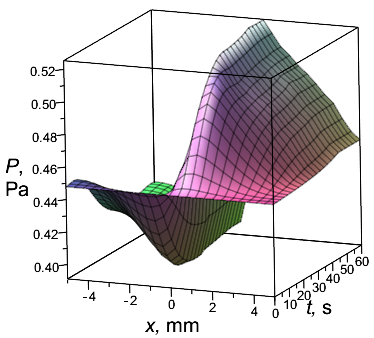}	\caption{\label{fig:PressureForCompositeSubstrateProblem}Spatiotemporal distribution of pressure exerted by the polymer mass on the substrate surface by gravity.}
\end{figure}

However, by the 55$^\mathrm{th}$ second, the temperature gradient changes direction starting to rise (Fig.~\ref{fig:TemperatureComparisonAndEvaporationRateForCompositeSubstrateProblem}a).
This is because the alcohol has almost completely evaporated around point $x=-X$ by this time, resulting in the vapor flux density $J\to 0$ (Fig.~\ref{fig:TemperatureComparisonAndEvaporationRateForCompositeSubstrateProblem}b).
The maximum temperature change of the film is about 6~K. In the case of the substrate, the maximum temperature change is approximately 2~K. Temperature dynamics affect the vapor flux density. At the process start ($t = 0$), the value $J\approx$ 2.05~g/(m$^2$s) is uniform along the free film surface (Fig.~\ref{fig:TemperatureComparisonAndEvaporationRateForCompositeSubstrateProblem}b), because the temperature $T_l$ is also distributed uniformly along the spatial coordinate $x$ (Fig.~\ref{fig:TemperatureEvolutionForCompositeSubstrateProblem}a). Later, the value $J$ decreases in the relatively cold region above the Teflon ($x<0$). At the process end, the vapor flux density also starts to decrease in the region of the aluminum ($x>0$) since the polymer mass fraction value $c$ approaches 1.

The redistribution of polymer mass $m_p$ can be inferred by the curve of pressure $P$ (Fig.~\ref{fig:PressureForCompositeSubstrateProblem}) exerted by this mass on the substrate by gravity ($P=\rho g h c \sim m_p$). The mass $m_p$ decreases over time above the Teflon layer and increases in the region of the aluminum surface. The reason for this is the convective mass transfer caused by the solutal Marangoni effect.

\begin{figure}
	\includegraphics[width=0.75\columnwidth]{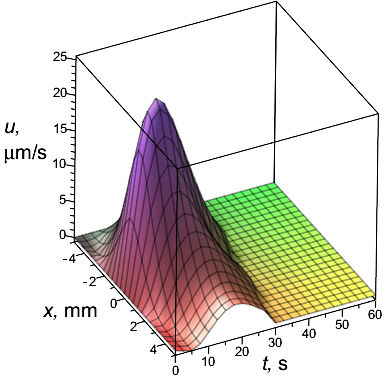} \caption{\label{fig:VelocityForCompositeSubstrateProblem} Spatiotemporal evolution of horizontal velocity averaged over the liquid layer height.}
\end{figure}

As can be seen from Fig.~\ref{fig:VelocityForCompositeSubstrateProblem} the flow velocity $u$ initially increases over time. The solution flow is directed along the positive direction of the $x$ axis, from the Teflon layer to the aluminum surface. The maximum value of $u$ is reached at time $t\approx$ 20~s in the point $x\approx 0 $ and is about 25~$\mu$m/s. As this indicates the horizontal flow velocity averaged over the liquid layer height it should be understood that the solutal Marangoni flow velocity on the free film surface will be greater than this value. Subsequently, the velocity decreases reaching the value $u=0$ by about the 35$^\mathrm{th}$ second i.e. the flow ceases. It follows that convective mass transfer occurs only during this period of time. At this point the solution concentration exceeds the critical value $c_g$ so the flow stops due to the high viscosity.

\begin{figure}
	\begin{minipage}[h]{0.9\linewidth}
		\center{\includegraphics[width=0.7\columnwidth]{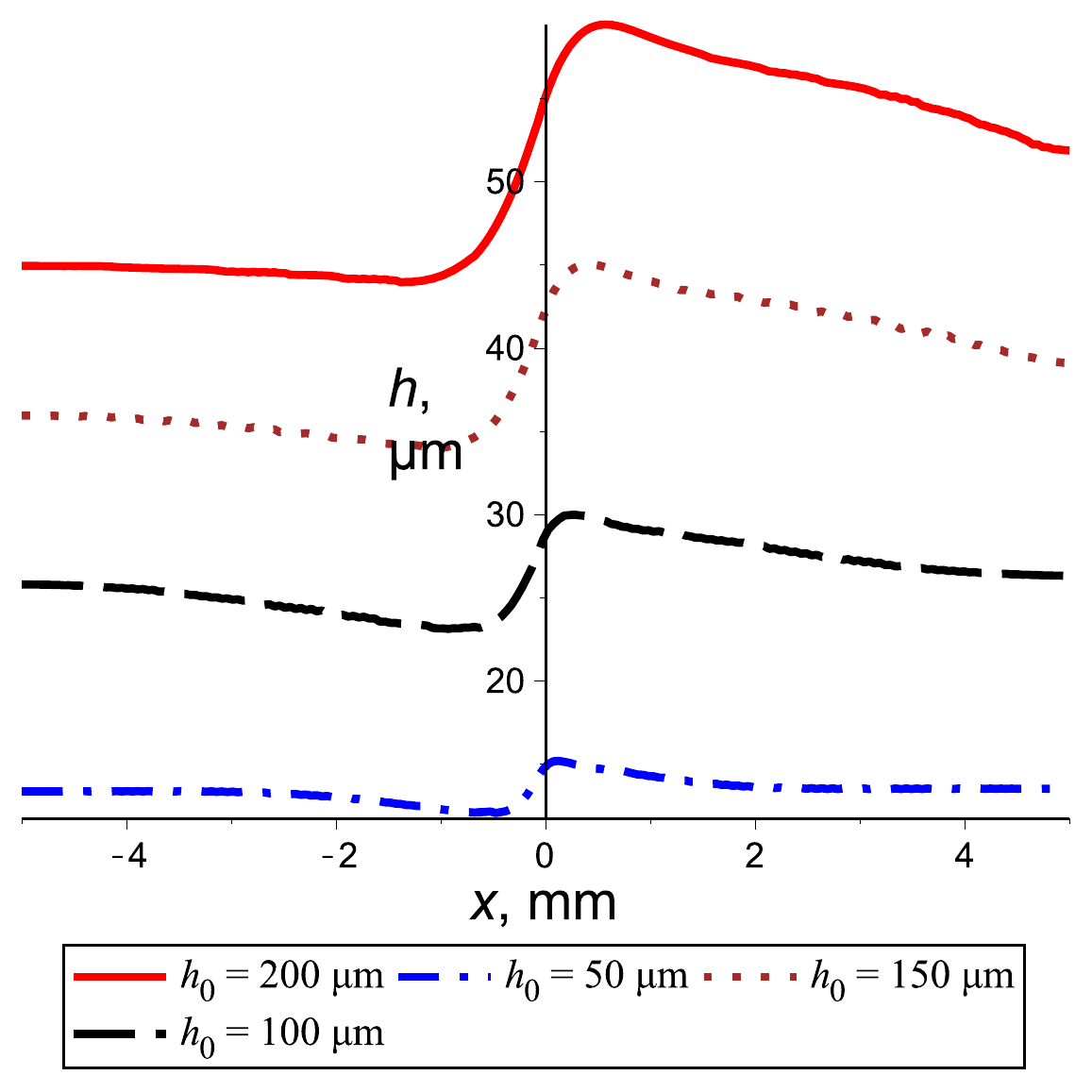} \\ (a)}
	\end{minipage}
	\\ \vspace{5 pt}
	\begin{minipage}[h]{0.9\linewidth}
		\center{\includegraphics[width=0.7\columnwidth]{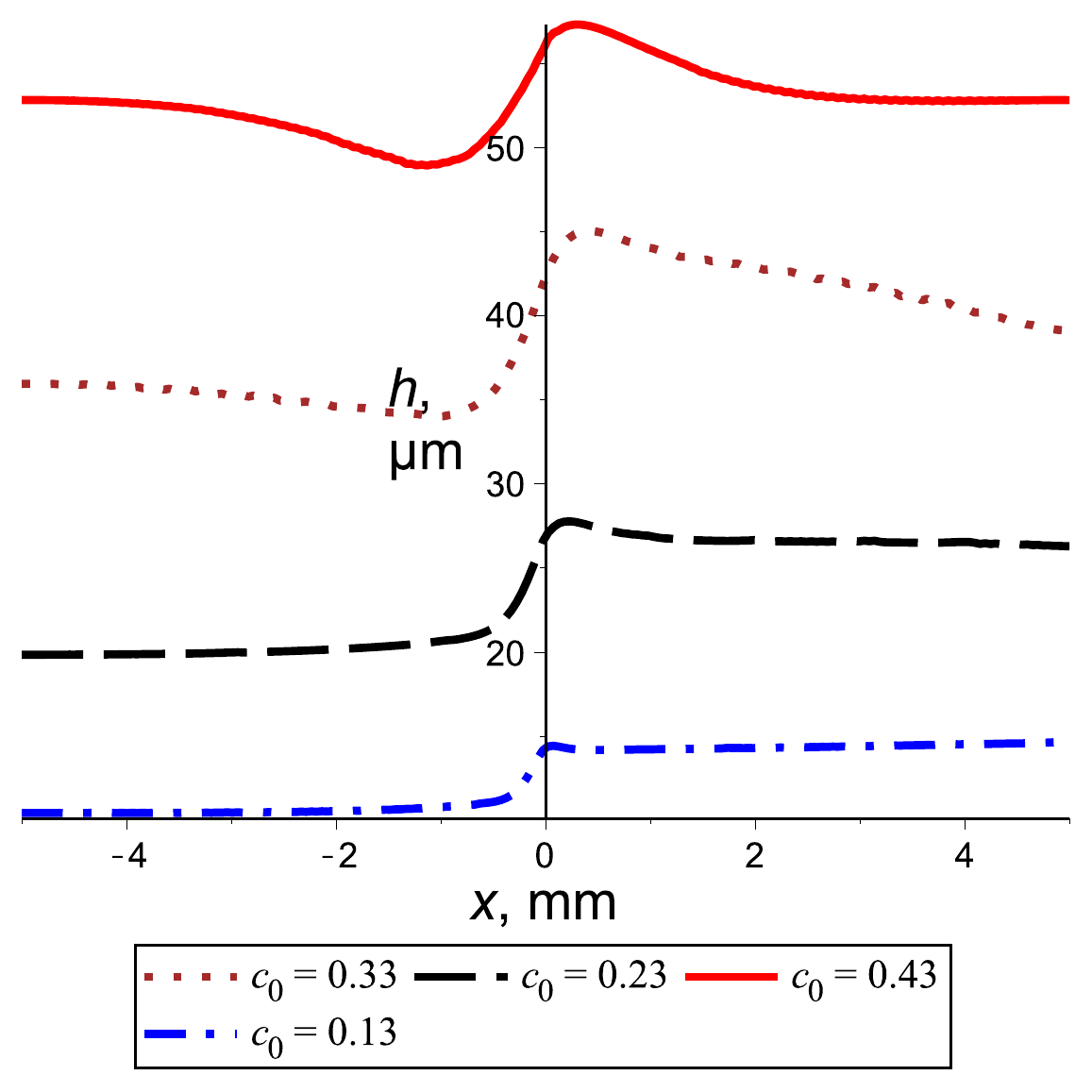} \\ (b)}
	\end{minipage}
	\caption{\label{fig:DependenceOfCoatingThicknessOnKeyParameters}Dependence of the thickness of the polymer coating on (a) the initial thickness of the liquid film and (b) the initial mass fraction of the solution.}
\end{figure}

To study the effect of key parameters on the final thickness of a relief polymer film, a series of numerical experiments was carried out. The thickness of the coating can be controlled by adjusting either the initial thickness of the liquid film (Fig.~\ref{fig:DependenceOfCoatingThicknessOnKeyParameters}a) or the concentration of the solution (Fig.~\ref{fig:DependenceOfCoatingThicknessOnKeyParameters}b). In all calculations, the value of $t_\mathrm{max}=$ 60~s has been used, as before. In only one instance, this time was not enough for complete evaporation of the liquid, when $h_0 =$ 200~$\mu$m (Fig.~\ref{fig:DependenceOfCoatingThicknessOnKeyParameters}a). In this case, the value of $t_\mathrm{max}=$ 100~s was used. The greater the values of $h_0$ and $c_0$, the thicker the polymer coating will become. In all cases, the shape of the relief coating remains unchanged. In some cases, the local curvature of this coating may vary slightly. The following empirical relationships have been identified based on the results: the average thickness of the polymer film approximately corresponds to values $h_\mathrm{ave}\approx h_0/a$ for $c_0=0.33$, and to $h_\mathrm{ave}\approx c_0 h_0 / (b -c_0)$ for $h_0=150$~$\mu$m. Here, $a$ and $b$ are the fitting parameters ($a\approx 3.73$, $b\approx 1.63$). These empirical dependencies can be used as a recipe for technological production. However, first, experimental validation is needed. It may be necessary to adjust the parameters $a$ and $b$ slightly, as the proposed model is a qualitative one. 

\section{Conclusion}
The results of the numerical computations indicate that solutal Marangoni flow plays a key role in the formation of a patterned polymer film as the  methanol is lost during drying. This confirms the hypothesis of the authors of the practical experiment~\cite{Cavadini2013}. The results obtained here are in qualitative agreement with their experimental data. This study allows us to better understand the mechanisms that can be used in evaporative lithography.

The rapid heat transfer from the aluminum surface to the liquid results in this region being less cold than the Teflon coating region, meaning that, above the aluminum surface, the vapor flux density is greater. Due to the relatively rapid evaporation, the polymer concentration in that local region increases more strongly resulting in a significant increase in the surface tension coefficient. Accordingly, the solutal Marangoni flow is directed towards the aluminum surface along the free surface of the liquid film. This transfers even more polymer there. As a consequence, this additionally promotes a local increase in the polymer concentration in the solution and intensification of fluid flow  affecting the admixture mass transfer. Thus, the Matthew effect emerges here. This flow increases until the viscosity begins to dominate. A further increase in viscosity due to an increase in the concentration of the solution leads to cessation of the flow.

The one-dimensional model described here is phenomenological in its nature. To obtain more accurate quantitative predictions a complex model should to be developed considering additional details such as the air flow velocity over the free film surface, $V_\mathrm{air}$, etc. In order to determine how the plate thickness $h_\mathrm{pl}$ affects the polymer film geometry a multidimensional model must be developed.

\begin{acknowledgments}
This work is supported by Grant No. 22-79-10216 from the Russian Science Foundation (https://rscf.ru/en/project/22-79-10216/).
\end{acknowledgments}

\section*{Data Availability Statement}

The data that support the findings of this study are
available from the corresponding author upon reasonable
request.

\appendix

\section{Estimation of several problem parameters}\label{secAppendix:parametersForCompositeSubstrateProblem}
The surface tension temperature gradient can be estimated using the empirical formula derived from the E{\"o}tv{\"o}s rule, $\sigma_T = -B (\rho_0 / M)^{2/3}$ where the constant $B\approx$ 2.1 g~cm$^2$/(s$^2$~K~mol$^{2/3}$). Substituting the values of the liquid parameters (methanol density $\rho_0 \approx$ 0.792~g/cm$^3$ and molar weight $M=$ 32~g/mol) results in $\sigma_T \approx -0.1782$ g/(K~s$^2$) and converting grams to kilograms provides the value $\sigma_T$, presented in Table~\ref{tab:ParametersInCompositeSubstrateProblem}.

\begin{figure}[h] 
	\includegraphics[width=0.8\columnwidth]{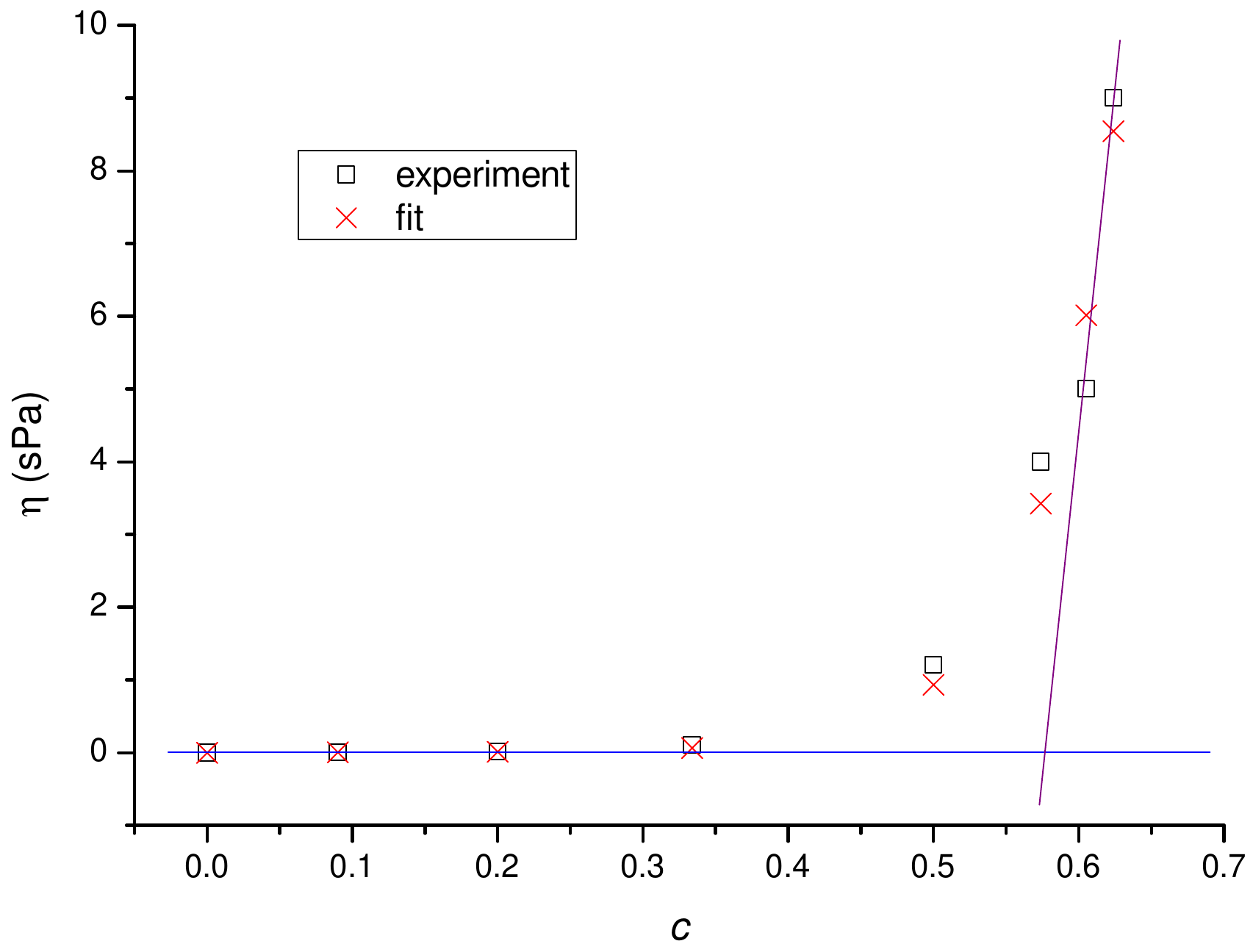} \caption{\label{fig:phaseTransitionConcentrationForCompositeSubstrateProblem}
Dependence of the solution  viscosity on the concentration at $T_l =$ 30~$^\circ$C.}
\end{figure}

Fig.~\ref{fig:phaseTransitionConcentrationForCompositeSubstrateProblem} shows the method for determination of the phase transition concentration $c_g$ (see Table~\ref{tab:ParametersInCompositeSubstrateProblem}). The experimental data are taken from the article~\cite{Toensmann2021}. Two main linear trends were determined, at the intersection of which the value $c_g\approx 0.58$ was obtained.

The convective heat exchange coefficients are calculated as follows: $\alpha_{sa}=\mathrm{Nu}_\mathrm{air} k_\mathrm{air}/ l_c$ and $\alpha_{ls} = \mathrm{Nu} k_l/ l_c$ (see Table~\ref{tab:ParametersInCompositeSubstrateProblem}, \ref{tab:DimensionlessParametersInCompositeSubstrateProblem}) where the Nusselt numbers $\mathrm{Nu}=C \mathrm{Ra}^n$ and $\mathrm{Nu}_\mathrm{air}=C \mathrm{Ra}_\mathrm{air}^n$~ \cite{Patochkina2016}. The values of the empirical parameters were taken from Table~1 of Ref.~\cite{Patochkina2016} ($C = 1.2\times 10^{-3}$, $n=0.9$). The value for $\alpha_{la}$ is taken from the article~\cite{Gatapova2014} in accordance with the value $\Delta T$ (see Table~\ref{tab:ParametersInCompositeSubstrateProblem}).

\section{Derivation of equations based on the mass conservation  law}\label{secAppendix:derivationOfEquationsForCompositeSubstrateProblem}

\begin{figure}[h]
	\includegraphics[width=0.5\columnwidth]{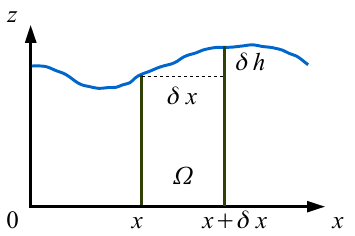} \caption{\label{fig:derivationOfEquationsForCompositeSubstrateProblem}
		Sketch for the derivation of equations based on the mass conservation law.}
\end{figure}

Consider the mass balance of the solution in an elementary volume $\Omega$ (Fig.~\ref{fig:derivationOfEquationsForCompositeSubstrateProblem}). The left border of $\Omega$ corresponds to the coordinate $x$, and the right border corresponds to the coordinate $x+\delta x$. The upper boundary ``air--liquid'' is determined by the coordinate $z=h$. The lower boundary ``liquid--substrate'' ($z=0$) is impermeable, thus there is no mass flux through it. The outflux of mass occurs through the upper boundary as a result of evaporation. In addition, the convective flow transfers the matter across the left and right boundaries of $\Omega$.

During $\Delta t$, the mass of the solution in the volume $\Omega$ will change due to changes in the volume itself and  in the solution  density,
\begin{equation}\label{eq:volumeChangeForCompositeSubstrateProblem}
	\Delta m = \delta y\, \delta x \left( h \rho|_{x,t+\Delta t} - h \rho|_{x,t} \right),
\end{equation}
where notation such as $h \rho|_{x,t}$ should be interpreted as $h(x,t) \rho(x,t)$. The mass flux density through the left and right (lateral) sides is written as $u\rho$. The length of the upper boundary is expressed as $$\sqrt{\delta x^2 +\delta h^2}=\delta x \sqrt{1+ \left( \frac{\delta h}{\delta x} \right)^2},$$ where $\delta h$ is the thickness  difference of the film on the left and right borders of $\Omega$.

Let us take into account the area of the lateral sides, $\delta y\, h$, and the area of the upper one, $\delta y\, \delta x \sqrt{1+(\delta h/\delta x)^2}$. The change in mass in $\Omega$ during $\Delta t$ is associated with the transfer of mass through the lateral and upper boundaries,
\begin{equation}\label{eq:massFlowForCompositeSubstrateProblem}
	\Delta m = -\Delta t\, \delta y\, \left( hu\rho|_{x+\delta x,t} - hu\rho|_{x,t} - J\, \delta x\, \sqrt{1+ \left( \frac{\delta h}{\delta x} \right)^2} \right).
\end{equation}
Equate the right parts of the expressions~\eqref{eq:volumeChangeForCompositeSubstrateProblem} and \eqref{eq:massFlowForCompositeSubstrateProblem}, divide by $\delta x\, \delta y\, \Delta t$, and get $$\frac{h \rho|_{x,t+\Delta t} - h \rho|_{x,t}}{\Delta t} = - \frac{hu\rho|_{x+\delta x,t} - hu\rho|_{x,t}}{\delta x} -J  \sqrt{1+ \left( \frac{\delta h}{\delta x} \right)^2}.$$
As a result of the limit transition ($\delta x \to 0$, $\Delta t \to 0$), we come to the equation
\begin{equation}\label{eq:dimensionalSolutionConservationLawForCompositeSubstrateProblem}
	\frac{\partial (h \rho)}{\partial t} + \frac{\partial (hu\rho)}{\partial x} = -J  \sqrt{1+ \left( \frac{\partial h}{\partial x} \right)^2},
\end{equation}
it follows from the lubrication approximation that $$\sqrt{1+ \left( \frac{\partial h}{\partial x} \right)^2}\approx 1.$$ In dimensionless form, the equation~\eqref{eq:dimensionalSolutionConservationLawForCompositeSubstrateProblem} takes the form~\eqref{eq:SolutionConservationLawForCompositeSubstrateProblem}.

Now let us consider the balance of the polymer mass in the elementary volume~$\Omega$. During the time $\Delta t$, the mass of the polymer changes by
\begin{equation}\label{eq:polimerMassChangeForCompositeSubstrateProblem}
	\Delta m_p = \delta x\, \delta y \left( h c \rho|_{x,t+\Delta t} - h c \rho|_{x,t} \right).
\end{equation}
This mass change is caused by convective and diffusive transport across the lateral boundaries of~$\Omega$,
\begin{equation}\label{eq:polimerMassFluxForCompositeSubstrateProblem}
	\Delta m_p = \Delta t\, \delta y \left( h c \rho u |_{x,t} - h c \rho u |_{x+\delta x,t} + h \xi |_{x,t} - h \xi |_{x+\delta x,t} \right),
\end{equation}
where the diffusion flux density of the admixture is $\xi = -D \rho\, \partial c/ \partial x$, this following from Fick's law. Equate the right parts of~\eqref{eq:polimerMassChangeForCompositeSubstrateProblem} and \eqref{eq:polimerMassFluxForCompositeSubstrateProblem}, divide them by $\delta x \, \delta y \, \Delta t$, and, as a result of the limiting transition, we obtain the convection--diffusion equation $$\frac{\partial (h c \rho)}{\partial t} = - \frac{\partial (h c \rho u)}{\partial x} - \frac{\partial (h \xi)}{\partial x}.$$ Next, this equation is rewritten in the following form,
$$h \rho \frac{\partial c}{\partial t} + c \frac{\partial (h \rho)}{\partial t} + c \frac{\partial (h \rho u)}{\partial x} +  h \rho u \frac{\partial c}{\partial x} = \frac{\partial}{\partial x} \left( h D \rho \frac{\partial c}{\partial x} \right).$$ Divide the left and right sides of the equation by $h \rho$ and taking into account~\eqref{eq:dimensionalSolutionConservationLawForCompositeSubstrateProblem} we get $$\frac{\partial c}{\partial t} + u \frac{\partial c}{\partial x} = \frac{1}{h \rho}  \frac{\partial}{\partial x} \left( D h \rho \frac{\partial c}{\partial x} \right) + \frac{J c}{h \rho} .$$ In dimensionless form, this equation can be written as~\eqref{eq:ConvectionDiffusionEqForCompositeSubstrateProblem}.

\nocite{*}
\bibliography{KolegovPoF2024}

\begin{thebibliography}{38}%
\makeatletter
\providecommand \@ifxundefined [1]{%
 \@ifx{#1\undefined}
}%
\providecommand \@ifnum [1]{%
 \ifnum #1\expandafter \@firstoftwo
 \else \expandafter \@secondoftwo
 \fi
}%
\providecommand \@ifx [1]{%
 \ifx #1\expandafter \@firstoftwo
 \else \expandafter \@secondoftwo
 \fi
}%
\providecommand \natexlab [1]{#1}%
\providecommand \enquote  [1]{``#1''}%
\providecommand \bibnamefont  [1]{#1}%
\providecommand \bibfnamefont [1]{#1}%
\providecommand \citenamefont [1]{#1}%
\providecommand \href@noop [0]{\@secondoftwo}%
\providecommand \href [0]{\begingroup \@sanitize@url \@href}%
\providecommand \@href[1]{\@@startlink{#1}\@@href}%
\providecommand \@@href[1]{\endgroup#1\@@endlink}%
\providecommand \@sanitize@url [0]{\catcode `\\12\catcode `\$12\catcode
  `\&12\catcode `\#12\catcode `\^12\catcode `\_12\catcode `\%12\relax}%
\providecommand \@@startlink[1]{}%
\providecommand \@@endlink[0]{}%
\providecommand \url  [0]{\begingroup\@sanitize@url \@url }%
\providecommand \@url [1]{\endgroup\@href {#1}{\urlprefix }}%
\providecommand \urlprefix  [0]{URL }%
\providecommand \Eprint [0]{\href }%
\providecommand \doibase [0]{https://doi.org/}%
\providecommand \selectlanguage [0]{\@gobble}%
\providecommand \bibinfo  [0]{\@secondoftwo}%
\providecommand \bibfield  [0]{\@secondoftwo}%
\providecommand \translation [1]{[#1]}%
\providecommand \BibitemOpen [0]{}%
\providecommand \bibitemStop [0]{}%
\providecommand \bibitemNoStop [0]{.\EOS\space}%
\providecommand \EOS [0]{\spacefactor3000\relax}%
\providecommand \BibitemShut  [1]{\csname bibitem#1\endcsname}%
\let\auto@bib@innerbib\@empty
\bibitem [{\citenamefont {Xue}\ \emph {et~al.}(2023)\citenamefont {Xue},
  \citenamefont {Wang}, \citenamefont {Chen},\ and\ \citenamefont
  {Liu}}]{Xue2023}%
  \BibitemOpen
  \bibfield  {author} {\bibinfo {author} {\bibfnamefont {W.}~\bibnamefont
  {Xue}}, \bibinfo {author} {\bibfnamefont {Y.}~\bibnamefont {Wang}}, \bibinfo
  {author} {\bibfnamefont {Z.}~\bibnamefont {Chen}},\ and\ \bibinfo {author}
  {\bibfnamefont {H.}~\bibnamefont {Liu}},\ }\bibfield  {title} {\enquote
  {\bibinfo {title} {An integrated model with stable numerical methods for
  fractured underground gas storage},}\ }\href
  {https://doi.org/10.1016/j.jclepro.2023.136268} {\bibfield  {journal}
  {\bibinfo  {journal} {Journal of Cleaner Production}\ }\textbf {\bibinfo
  {volume} {393}},\ \bibinfo {pages} {136268} (\bibinfo {year}
  {2023})}\BibitemShut {NoStop}%
\bibitem [{\citenamefont {Xue}\ \emph {et~al.}(2024)\citenamefont {Xue},
  \citenamefont {Wang}, \citenamefont {Liang}, \citenamefont {Wang},\ and\
  \citenamefont {Ren}}]{Xue2024}%
  \BibitemOpen
  \bibfield  {author} {\bibinfo {author} {\bibfnamefont {W.}~\bibnamefont
  {Xue}}, \bibinfo {author} {\bibfnamefont {Y.}~\bibnamefont {Wang}}, \bibinfo
  {author} {\bibfnamefont {Y.}~\bibnamefont {Liang}}, \bibinfo {author}
  {\bibfnamefont {T.}~\bibnamefont {Wang}},\ and\ \bibinfo {author}
  {\bibfnamefont {B.}~\bibnamefont {Ren}},\ }\bibfield  {title} {\enquote
  {\bibinfo {title} {Efficient hydraulic and thermal simulation model of the
  multi-phase natural gas production system with variable speed compressors},}\
  }\href {https://doi.org/10.1016/j.applthermaleng.2024.122411} {\bibfield
  {journal} {\bibinfo  {journal} {Applied Thermal Engineering}\ }\textbf
  {\bibinfo {volume} {242}},\ \bibinfo {pages} {122411} (\bibinfo {year}
  {2024})}\BibitemShut {NoStop}%
\bibitem [{\citenamefont {Deegan}\ \emph {et~al.}(1997)\citenamefont {Deegan},
  \citenamefont {Bakajin}, \citenamefont {Dupont}, \citenamefont {Huber},
  \citenamefont {Nagel},\ and\ \citenamefont {Witten}}]{Deegan1997}%
  \BibitemOpen
  \bibfield  {author} {\bibinfo {author} {\bibfnamefont {R.~D.}\ \bibnamefont
  {Deegan}}, \bibinfo {author} {\bibfnamefont {O.}~\bibnamefont {Bakajin}},
  \bibinfo {author} {\bibfnamefont {T.~F.}\ \bibnamefont {Dupont}}, \bibinfo
  {author} {\bibfnamefont {G.}~\bibnamefont {Huber}}, \bibinfo {author}
  {\bibfnamefont {S.~R.}\ \bibnamefont {Nagel}},\ and\ \bibinfo {author}
  {\bibfnamefont {T.~A.}\ \bibnamefont {Witten}},\ }\bibfield  {title}
  {\enquote {\bibinfo {title} {Capillary flow as the cause of ring stains from
  dried liquid drops},}\ }\href {https://doi.org/10.1038/39827} {\bibfield
  {journal} {\bibinfo  {journal} {Nature}\ }\textbf {\bibinfo {volume} {389}},\
  \bibinfo {pages} {827--829} (\bibinfo {year} {1997})}\BibitemShut {NoStop}%
\bibitem [{\citenamefont {Deegan}(2000)}]{Deegan2000475}%
  \BibitemOpen
  \bibfield  {author} {\bibinfo {author} {\bibfnamefont {R.~D.}\ \bibnamefont
  {Deegan}},\ }\bibfield  {title} {\enquote {\bibinfo {title} {Pattern
  formation in drying drops},}\ }\href
  {https://doi.org/10.1103/physreve.61.475} {\bibfield  {journal} {\bibinfo
  {journal} {Physical Review E}\ }\textbf {\bibinfo {volume} {61}},\ \bibinfo
  {pages} {475--485} (\bibinfo {year} {2000})}\BibitemShut {NoStop}%
\bibitem [{\citenamefont {Sefiane}(2014)}]{Sefiane2014}%
  \BibitemOpen
  \bibfield  {author} {\bibinfo {author} {\bibfnamefont {K.}~\bibnamefont
  {Sefiane}},\ }\bibfield  {title} {\enquote {\bibinfo {title} {Patterns from
  drying drops},}\ }\href {https://doi.org/10.1016/j.cis.2013.05.002}
  {\bibfield  {journal} {\bibinfo  {journal} {Advances in Colloid and Interface
  Science}\ }\textbf {\bibinfo {volume} {206}},\ \bibinfo {pages} {372--381}
  (\bibinfo {year} {2014})}\BibitemShut {NoStop}%
\bibitem [{\citenamefont {Routh}\ and\ \citenamefont
  {Russel}(1998)}]{Routh1998}%
  \BibitemOpen
  \bibfield  {author} {\bibinfo {author} {\bibfnamefont {A.~F.}\ \bibnamefont
  {Routh}}\ and\ \bibinfo {author} {\bibfnamefont {W.~B.}\ \bibnamefont
  {Russel}},\ }\bibfield  {title} {\enquote {\bibinfo {title} {Horizontal
  drying fronts during solvent evaporation from latex films},}\ }\href
  {https://doi.org/10.1002/aic.690440916} {\bibfield  {journal} {\bibinfo
  {journal} {{AIChE} Journal}\ }\textbf {\bibinfo {volume} {44}},\ \bibinfo
  {pages} {2088--2098} (\bibinfo {year} {1998})}\BibitemShut {NoStop}%
\bibitem [{\citenamefont {Harris}\ \emph {et~al.}(2007)\citenamefont {Harris},
  \citenamefont {Hu}, \citenamefont {Conrad},\ and\ \citenamefont
  {Lewis}}]{Harris2007}%
  \BibitemOpen
  \bibfield  {author} {\bibinfo {author} {\bibfnamefont {D.~J.}\ \bibnamefont
  {Harris}}, \bibinfo {author} {\bibfnamefont {H.}~\bibnamefont {Hu}}, \bibinfo
  {author} {\bibfnamefont {J.~C.}\ \bibnamefont {Conrad}},\ and\ \bibinfo
  {author} {\bibfnamefont {J.~A.}\ \bibnamefont {Lewis}},\ }\bibfield  {title}
  {\enquote {\bibinfo {title} {Patterning colloidal films via evaporative
  lithography},}\ }\href {https://doi.org/10.1103/physrevlett.98.148301}
  {\bibfield  {journal} {\bibinfo  {journal} {Physical Review Letters}\
  }\textbf {\bibinfo {volume} {98}},\ \bibinfo {pages} {148301} (\bibinfo
  {year} {2007})}\BibitemShut {NoStop}%
\bibitem [{\citenamefont {Harris}\ and\ \citenamefont
  {Lewis}(2008)}]{Harris2008}%
  \BibitemOpen
  \bibfield  {author} {\bibinfo {author} {\bibfnamefont {D.~J.}\ \bibnamefont
  {Harris}}\ and\ \bibinfo {author} {\bibfnamefont {J.~A.}\ \bibnamefont
  {Lewis}},\ }\bibfield  {title} {\enquote {\bibinfo {title} {Marangoni effects
  on evaporative lithographic patterning of colloidal films},}\ }\href
  {https://doi.org/10.1021/la8000637} {\bibfield  {journal} {\bibinfo
  {journal} {Langmuir}\ }\textbf {\bibinfo {volume} {24}},\ \bibinfo {pages}
  {3681--3685} (\bibinfo {year} {2008})}\BibitemShut {NoStop}%
\bibitem [{\citenamefont {Kolegov}\ and\ \citenamefont
  {Barash}(2020)}]{Kolegov2020}%
  \BibitemOpen
  \bibfield  {author} {\bibinfo {author} {\bibfnamefont {K.~S.}\ \bibnamefont
  {Kolegov}}\ and\ \bibinfo {author} {\bibfnamefont {L.~Y.}\ \bibnamefont
  {Barash}},\ }\bibfield  {title} {\enquote {\bibinfo {title} {Applying
  droplets and films in evaporative lithography},}\ }\href
  {https://doi.org/10.1016/j.cis.2020.102271} {\bibfield  {journal} {\bibinfo
  {journal} {Advances in Colloid and Interface Science}\ }\textbf {\bibinfo
  {volume} {285}},\ \bibinfo {pages} {102271} (\bibinfo {year}
  {2020})}\BibitemShut {NoStop}%
\bibitem [{\citenamefont {D{\textquotesingle}Ambrosio}\ \emph
  {et~al.}(2023)\citenamefont {D{\textquotesingle}Ambrosio}, \citenamefont
  {Wilson}, \citenamefont {Wray},\ and\ \citenamefont {Duffy}}]{Ambrosio2023}%
  \BibitemOpen
  \bibfield  {author} {\bibinfo {author} {\bibfnamefont {H.-M.}\ \bibnamefont
  {D{\textquotesingle}Ambrosio}}, \bibinfo {author} {\bibfnamefont {S.~K.}\
  \bibnamefont {Wilson}}, \bibinfo {author} {\bibfnamefont {A.~W.}\
  \bibnamefont {Wray}},\ and\ \bibinfo {author} {\bibfnamefont {B.~R.}\
  \bibnamefont {Duffy}},\ }\bibfield  {title} {\enquote {\bibinfo {title} {The
  effect of the spatial variation of the evaporative flux on the deposition
  from a thin sessile droplet},}\ }\href {https://doi.org/10.1017/jfm.2023.503}
  {\bibfield  {journal} {\bibinfo  {journal} {Journal of Fluid Mechanics}\
  }\textbf {\bibinfo {volume} {970}} (\bibinfo {year} {2023}),\
  10.1017/jfm.2023.503}\BibitemShut {NoStop}%
\bibitem [{\citenamefont {Li}, \citenamefont {Kar},\ and\ \citenamefont
  {Kumar}(2019)}]{Li201972}%
  \BibitemOpen
  \bibfield  {author} {\bibinfo {author} {\bibfnamefont {T.}~\bibnamefont
  {Li}}, \bibinfo {author} {\bibfnamefont {A.}~\bibnamefont {Kar}},\ and\
  \bibinfo {author} {\bibfnamefont {R.}~\bibnamefont {Kumar}},\ }\bibfield
  {title} {\enquote {\bibinfo {title} {Marangoni circulation by {UV} light
  modulation on sessile drop for particle agglomeration},}\ }\href
  {https://doi.org/10.1017/jfm.2019.373} {\bibfield  {journal} {\bibinfo
  {journal} {Journal of Fluid Mechanics}\ }\textbf {\bibinfo {volume} {873}},\
  \bibinfo {pages} {72--88} (\bibinfo {year} {2019})}\BibitemShut {NoStop}%
\bibitem [{\citenamefont {Al-Muzaiqer}\ \emph
  {et~al.}(2021{\natexlab{a}})\citenamefont {Al-Muzaiqer}, \citenamefont
  {Ivanova}, \citenamefont {Fliagin},\ and\ \citenamefont
  {Lebedev-Stepanov}}]{AlMuzaiqer2021126550}%
  \BibitemOpen
  \bibfield  {author} {\bibinfo {author} {\bibfnamefont {M.~A.}\ \bibnamefont
  {Al-Muzaiqer}}, \bibinfo {author} {\bibfnamefont {N.~A.}\ \bibnamefont
  {Ivanova}}, \bibinfo {author} {\bibfnamefont {V.~M.}\ \bibnamefont
  {Fliagin}},\ and\ \bibinfo {author} {\bibfnamefont {P.~V.}\ \bibnamefont
  {Lebedev-Stepanov}},\ }\bibfield  {title} {\enquote {\bibinfo {title}
  {Transport and assembling microparticles via {M}arangoni flows in heating and
  cooling modes},}\ }\href {https://doi.org/10.1016/j.colsurfa.2021.126550}
  {\bibfield  {journal} {\bibinfo  {journal} {Colloids and Surfaces A:
  Physicochemical and Engineering Aspects}\ }\textbf {\bibinfo {volume}
  {621}},\ \bibinfo {pages} {126550} (\bibinfo {year}
  {2021}{\natexlab{a}})}\BibitemShut {NoStop}%
\bibitem [{\citenamefont {Al-Muzaiqer}\ \emph
  {et~al.}(2021{\natexlab{b}})\citenamefont {Al-Muzaiqer}, \citenamefont
  {Kolegov}, \citenamefont {Ivanova},\ and\ \citenamefont
  {Fliagin}}]{AlMuzaiqer2021}%
  \BibitemOpen
  \bibfield  {author} {\bibinfo {author} {\bibfnamefont {M.~A.}\ \bibnamefont
  {Al-Muzaiqer}}, \bibinfo {author} {\bibfnamefont {K.~S.}\ \bibnamefont
  {Kolegov}}, \bibinfo {author} {\bibfnamefont {N.~A.}\ \bibnamefont
  {Ivanova}},\ and\ \bibinfo {author} {\bibfnamefont {V.~M.}\ \bibnamefont
  {Fliagin}},\ }\bibfield  {title} {\enquote {\bibinfo {title} {Nonuniform
  heating of a substrate in evaporative lithography},}\ }\href
  {https://doi.org/10.1063/5.0061713} {\bibfield  {journal} {\bibinfo
  {journal} {Physics of Fluids}\ }\textbf {\bibinfo {volume} {33}},\ \bibinfo
  {pages} {092101} (\bibinfo {year} {2021}{\natexlab{b}})}\BibitemShut
  {NoStop}%
\bibitem [{\citenamefont {Farzeena}\ and\ \citenamefont
  {Varanakkottu}(2022)}]{Farzeena2022}%
  \BibitemOpen
  \bibfield  {author} {\bibinfo {author} {\bibfnamefont {C.}~\bibnamefont
  {Farzeena}}\ and\ \bibinfo {author} {\bibfnamefont {S.~N.}\ \bibnamefont
  {Varanakkottu}},\ }\bibfield  {title} {\enquote {\bibinfo {title} {Patterning
  of metallic nanoparticles over solid surfaces from sessile droplets by
  thermoplasmonically controlled liquid flow},}\ }\href
  {https://doi.org/10.1021/acs.langmuir.1c02739} {\bibfield  {journal}
  {\bibinfo  {journal} {Langmuir}\ }\textbf {\bibinfo {volume} {38}},\ \bibinfo
  {pages} {2003--2013} (\bibinfo {year} {2022})}\BibitemShut {NoStop}%
\bibitem [{\citenamefont {Perkins-Howard}\ \emph {et~al.}(2022)\citenamefont
  {Perkins-Howard}, \citenamefont {Walker}, \citenamefont {Do}, \citenamefont
  {Senadheera}, \citenamefont {Hazzazi}, \citenamefont {Grundhoefer},
  \citenamefont {Daniels-Race},\ and\ \citenamefont
  {Garno}}]{PerkinsHoward2022}%
  \BibitemOpen
  \bibfield  {author} {\bibinfo {author} {\bibfnamefont {B.}~\bibnamefont
  {Perkins-Howard}}, \bibinfo {author} {\bibfnamefont {A.~R.}\ \bibnamefont
  {Walker}}, \bibinfo {author} {\bibfnamefont {Q.}~\bibnamefont {Do}}, \bibinfo
  {author} {\bibfnamefont {D.~I.}\ \bibnamefont {Senadheera}}, \bibinfo
  {author} {\bibfnamefont {F.}~\bibnamefont {Hazzazi}}, \bibinfo {author}
  {\bibfnamefont {J.~P.}\ \bibnamefont {Grundhoefer}}, \bibinfo {author}
  {\bibfnamefont {T.}~\bibnamefont {Daniels-Race}},\ and\ \bibinfo {author}
  {\bibfnamefont {J.~C.}\ \bibnamefont {Garno}},\ }\bibfield  {title} {\enquote
  {\bibinfo {title} {Surface wettability drives the crystalline surface
  assembly of monodisperse spheres in evaporative colloidal lithography},}\
  }\href {https://doi.org/10.1021/acs.jpcc.1c07098} {\bibfield  {journal}
  {\bibinfo  {journal} {The Journal of Physical Chemistry C}\ }\textbf
  {\bibinfo {volume} {126}},\ \bibinfo {pages} {505--516} (\bibinfo {year}
  {2022})}\BibitemShut {NoStop}%
\bibitem [{\citenamefont {Hegde}\ \emph {et~al.}(2022)\citenamefont {Hegde},
  \citenamefont {Chatterjee}, \citenamefont {Rasheed}, \citenamefont
  {Chakravortty},\ and\ \citenamefont {Basu}}]{Hegde2022}%
  \BibitemOpen
  \bibfield  {author} {\bibinfo {author} {\bibfnamefont {O.}~\bibnamefont
  {Hegde}}, \bibinfo {author} {\bibfnamefont {R.}~\bibnamefont {Chatterjee}},
  \bibinfo {author} {\bibfnamefont {A.}~\bibnamefont {Rasheed}}, \bibinfo
  {author} {\bibfnamefont {D.}~\bibnamefont {Chakravortty}},\ and\ \bibinfo
  {author} {\bibfnamefont {S.}~\bibnamefont {Basu}},\ }\bibfield  {title}
  {\enquote {\bibinfo {title} {Multiscale vapor-mediated dendritic pattern
  formation and bacterial aggregation in complex respiratory biofluid
  droplets},}\ }\href {https://doi.org/10.1016/j.jcis.2021.09.158} {\bibfield
  {journal} {\bibinfo  {journal} {Journal of Colloid and Interface Science}\
  }\textbf {\bibinfo {volume} {606}},\ \bibinfo {pages} {2011--2023} (\bibinfo
  {year} {2022})}\BibitemShut {NoStop}%
\bibitem [{\citenamefont {Li}\ \emph {et~al.}(2021)\citenamefont {Li},
  \citenamefont {Chen}, \citenamefont {Zhu}, \citenamefont {Liao},
  \citenamefont {Ye}, \citenamefont {Yang}, \citenamefont {Li}, \citenamefont
  {Li},\ and\ \citenamefont {Yang}}]{Li2021a}%
  \BibitemOpen
  \bibfield  {author} {\bibinfo {author} {\bibfnamefont {D.}~\bibnamefont
  {Li}}, \bibinfo {author} {\bibfnamefont {R.}~\bibnamefont {Chen}}, \bibinfo
  {author} {\bibfnamefont {X.}~\bibnamefont {Zhu}}, \bibinfo {author}
  {\bibfnamefont {Q.}~\bibnamefont {Liao}}, \bibinfo {author} {\bibfnamefont
  {D.}~\bibnamefont {Ye}}, \bibinfo {author} {\bibfnamefont {Y.}~\bibnamefont
  {Yang}}, \bibinfo {author} {\bibfnamefont {W.}~\bibnamefont {Li}}, \bibinfo
  {author} {\bibfnamefont {H.}~\bibnamefont {Li}},\ and\ \bibinfo {author}
  {\bibfnamefont {Y.}~\bibnamefont {Yang}},\ }\bibfield  {title} {\enquote
  {\bibinfo {title} {Light-fueled beating coffee-ring deposition for droplet
  evaporative crystallization},}\ }\href
  {https://doi.org/10.1021/acs.analchem.1c00605} {\bibfield  {journal}
  {\bibinfo  {journal} {Analytical Chemistry}\ }\textbf {\bibinfo {volume}
  {93}},\ \bibinfo {pages} {8817--8825} (\bibinfo {year} {2021})}\BibitemShut
  {NoStop}%
\bibitem [{\citenamefont {Goy}\ \emph {et~al.}(2022)\citenamefont {Goy},
  \citenamefont {Bruni}, \citenamefont {Girot}, \citenamefont {Delville},\ and\
  \citenamefont {Delabre}}]{Goy2022}%
  \BibitemOpen
  \bibfield  {author} {\bibinfo {author} {\bibfnamefont {N.-A.}\ \bibnamefont
  {Goy}}, \bibinfo {author} {\bibfnamefont {N.}~\bibnamefont {Bruni}}, \bibinfo
  {author} {\bibfnamefont {A.}~\bibnamefont {Girot}}, \bibinfo {author}
  {\bibfnamefont {J.-P.}\ \bibnamefont {Delville}},\ and\ \bibinfo {author}
  {\bibfnamefont {U.}~\bibnamefont {Delabre}},\ }\bibfield  {title} {\enquote
  {\bibinfo {title} {Thermal {M}arangoni trapping driven by laser absorption in
  evaporating droplets for particle deposition},}\ }\href
  {https://doi.org/10.1039/d2sm01019d} {\bibfield  {journal} {\bibinfo
  {journal} {Soft Matter}\ }\textbf {\bibinfo {volume} {18}},\ \bibinfo {pages}
  {7949--7958} (\bibinfo {year} {2022})}\BibitemShut {NoStop}%
\bibitem [{\citenamefont {Berneman}\ \emph {et~al.}(2021)\citenamefont
  {Berneman}, \citenamefont {Jimidar}, \citenamefont {Geite}, \citenamefont
  {Gardeniers},\ and\ \citenamefont {Desmet}}]{Berneman2021}%
  \BibitemOpen
  \bibfield  {author} {\bibinfo {author} {\bibfnamefont {N.}~\bibnamefont
  {Berneman}}, \bibinfo {author} {\bibfnamefont {I.}~\bibnamefont {Jimidar}},
  \bibinfo {author} {\bibfnamefont {W.~V.}\ \bibnamefont {Geite}}, \bibinfo
  {author} {\bibfnamefont {H.}~\bibnamefont {Gardeniers}},\ and\ \bibinfo
  {author} {\bibfnamefont {G.}~\bibnamefont {Desmet}},\ }\bibfield  {title}
  {\enquote {\bibinfo {title} {Rapid vacuum-driven monolayer assembly of
  microparticles on the surface of perforated microfluidic devices},}\ }\href
  {https://doi.org/10.1016/j.powtec.2021.05.079} {\bibfield  {journal}
  {\bibinfo  {journal} {Powder Technology}\ }\textbf {\bibinfo {volume}
  {390}},\ \bibinfo {pages} {330--338} (\bibinfo {year} {2021})}\BibitemShut
  {NoStop}%
\bibitem [{\citenamefont {Corletto}\ and\ \citenamefont
  {Shapter}(2021)}]{Corletto2021}%
  \BibitemOpen
  \bibfield  {author} {\bibinfo {author} {\bibfnamefont {A.}~\bibnamefont
  {Corletto}}\ and\ \bibinfo {author} {\bibfnamefont {J.~G.}\ \bibnamefont
  {Shapter}},\ }\bibfield  {title} {\enquote {\bibinfo {title}
  {Thickness/morphology of functional material patterned by topographical
  discontinuous dewetting},}\ }\href {https://doi.org/10.1002/nano.202000301}
  {\bibfield  {journal} {\bibinfo  {journal} {Nano Select}\ }\textbf {\bibinfo
  {volume} {2}},\ \bibinfo {pages} {1723--1740} (\bibinfo {year}
  {2021})}\BibitemShut {NoStop}%
\bibitem [{\citenamefont {Gambaryan-Roisman}(2010)}]{GambaryanRoisman2010}%
  \BibitemOpen
  \bibfield  {author} {\bibinfo {author} {\bibfnamefont {T.}~\bibnamefont
  {Gambaryan-Roisman}},\ }\bibfield  {title} {\enquote {\bibinfo {title}
  {Marangoni convection, evaporation and interface deformation in liquid films
  on heated substrates with non-uniform thermal conductivity},}\ }\href
  {https://doi.org/10.1016/j.ijheatmasstransfer.2009.09.017} {\bibfield
  {journal} {\bibinfo  {journal} {International Journal of Heat and Mass
  Transfer}\ }\textbf {\bibinfo {volume} {53}},\ \bibinfo {pages} {390--402}
  (\bibinfo {year} {2010})}\BibitemShut {NoStop}%
\bibitem [{\citenamefont {Gambaryan-Roisman}(2012)}]{GambaryanRoisman2012}%
  \BibitemOpen
  \bibfield  {author} {\bibinfo {author} {\bibfnamefont {T.}~\bibnamefont
  {Gambaryan-Roisman}},\ }\bibfield  {title} {\enquote {\bibinfo {title}
  {Marangoni-induced deformation of evaporating liquid films on composite
  substrates},}\ }\href {https://doi.org/10.1007/s10665-011-9498-9} {\bibfield
  {journal} {\bibinfo  {journal} {Journal of Engineering Mathematics}\ }\textbf
  {\bibinfo {volume} {73}},\ \bibinfo {pages} {39--52} (\bibinfo {year}
  {2012})}\BibitemShut {NoStop}%
\bibitem [{\citenamefont {Gambaryan-Roisman}(2015)}]{GambaryanRoisman2015}%
  \BibitemOpen
  \bibfield  {author} {\bibinfo {author} {\bibfnamefont {T.}~\bibnamefont
  {Gambaryan-Roisman}},\ }\bibfield  {title} {\enquote {\bibinfo {title}
  {Modulation of {M}arangoni convection in liquid films},}\ }\href
  {https://doi.org/10.1016/j.cis.2015.02.003} {\bibfield  {journal} {\bibinfo
  {journal} {Advances in Colloid and Interface Science}\ }\textbf {\bibinfo
  {volume} {222}},\ \bibinfo {pages} {319--331} (\bibinfo {year}
  {2015})}\BibitemShut {NoStop}%
\bibitem [{\citenamefont {Iqbal}\ \emph {et~al.}(2022)\citenamefont {Iqbal},
  \citenamefont {Matsumoto}, \citenamefont {Carlson}, \citenamefont {Peters},
  \citenamefont {Funari}, \citenamefont {Sen},\ and\ \citenamefont
  {Shen}}]{Iqbal2022}%
  \BibitemOpen
  \bibfield  {author} {\bibinfo {author} {\bibfnamefont {R.}~\bibnamefont
  {Iqbal}}, \bibinfo {author} {\bibfnamefont {A.}~\bibnamefont {Matsumoto}},
  \bibinfo {author} {\bibfnamefont {D.}~\bibnamefont {Carlson}}, \bibinfo
  {author} {\bibfnamefont {K.~T.}\ \bibnamefont {Peters}}, \bibinfo {author}
  {\bibfnamefont {R.}~\bibnamefont {Funari}}, \bibinfo {author} {\bibfnamefont
  {A.~K.}\ \bibnamefont {Sen}},\ and\ \bibinfo {author} {\bibfnamefont {A.~Q.}\
  \bibnamefont {Shen}},\ }\bibfield  {title} {\enquote {\bibinfo {title}
  {Evaporation driven smart patterning of microparticles on a rigid-soft
  composite substrate},}\ }\href {https://doi.org/10.1016/j.jcis.2022.05.087}
  {\bibfield  {journal} {\bibinfo  {journal} {Journal of Colloid and Interface
  Science}\ }\textbf {\bibinfo {volume} {623}},\ \bibinfo {pages} {927--937}
  (\bibinfo {year} {2022})}\BibitemShut {NoStop}%
\bibitem [{\citenamefont {Goel}\ \emph {et~al.}(2019)\citenamefont {Goel},
  \citenamefont {Choudhury}, \citenamefont {Aqeel}, \citenamefont {Li},
  \citenamefont {Shao},\ and\ \citenamefont {Duan}}]{Goel2019}%
  \BibitemOpen
  \bibfield  {author} {\bibinfo {author} {\bibfnamefont {P.}~\bibnamefont
  {Goel}}, \bibinfo {author} {\bibfnamefont {M.~D.}\ \bibnamefont {Choudhury}},
  \bibinfo {author} {\bibfnamefont {A.~B.}\ \bibnamefont {Aqeel}}, \bibinfo
  {author} {\bibfnamefont {X.}~\bibnamefont {Li}}, \bibinfo {author}
  {\bibfnamefont {L.-H.}\ \bibnamefont {Shao}},\ and\ \bibinfo {author}
  {\bibfnamefont {H.}~\bibnamefont {Duan}},\ }\bibfield  {title} {\enquote
  {\bibinfo {title} {Effect of thermal conductivity on enhanced evaporation of
  water droplets from heated graphene{\textendash}{PDMS} composite surfaces},}\
  }\href {https://doi.org/10.1021/acs.langmuir.9b00799} {\bibfield  {journal}
  {\bibinfo  {journal} {Langmuir}\ }\textbf {\bibinfo {volume} {35}},\ \bibinfo
  {pages} {6916--6921} (\bibinfo {year} {2019})}\BibitemShut {NoStop}%
\bibitem [{\citenamefont {Zhao}, \citenamefont {Zhang},\ and\ \citenamefont
  {Si}(2023)}]{Zhao2023}%
  \BibitemOpen
  \bibfield  {author} {\bibinfo {author} {\bibfnamefont {C.}~\bibnamefont
  {Zhao}}, \bibinfo {author} {\bibfnamefont {Z.}~\bibnamefont {Zhang}},\ and\
  \bibinfo {author} {\bibfnamefont {T.}~\bibnamefont {Si}},\ }\bibfield
  {title} {\enquote {\bibinfo {title} {Fluctuation-driven instability of
  nanoscale liquid films on chemically heterogeneous substrates},}\ }\href
  {https://doi.org/10.1063/5.0159155} {\bibfield  {journal} {\bibinfo
  {journal} {Physics of Fluids}\ }\textbf {\bibinfo {volume} {35}} (\bibinfo
  {year} {2023}),\ 10.1063/5.0159155}\BibitemShut {NoStop}%
\bibitem [{\citenamefont {Abe}\ \emph {et~al.}(2024)\citenamefont {Abe},
  \citenamefont {Atkinson}, \citenamefont {Cheung}, \citenamefont {Liang},
  \citenamefont {Goehring},\ and\ \citenamefont {Inasawa}}]{Abe2024}%
  \BibitemOpen
  \bibfield  {author} {\bibinfo {author} {\bibfnamefont {K.}~\bibnamefont
  {Abe}}, \bibinfo {author} {\bibfnamefont {P.~S.}\ \bibnamefont {Atkinson}},
  \bibinfo {author} {\bibfnamefont {C.~S.}\ \bibnamefont {Cheung}}, \bibinfo
  {author} {\bibfnamefont {H.}~\bibnamefont {Liang}}, \bibinfo {author}
  {\bibfnamefont {L.}~\bibnamefont {Goehring}},\ and\ \bibinfo {author}
  {\bibfnamefont {S.}~\bibnamefont {Inasawa}},\ }\bibfield  {title} {\enquote
  {\bibinfo {title} {Dynamics of drying colloidal suspensions, measured by
  optical coherence tomography},}\ }\href {https://doi.org/10.1039/d3sm01560b}
  {\bibfield  {journal} {\bibinfo  {journal} {Soft Matter}\ } (\bibinfo {year}
  {2024}),\ 10.1039/d3sm01560b}\BibitemShut {NoStop}%
\bibitem [{\citenamefont {Cavadini}\ \emph {et~al.}(2013)\citenamefont
  {Cavadini}, \citenamefont {Krenn}, \citenamefont {Scharfer},\ and\
  \citenamefont {Schabel}}]{Cavadini2013}%
  \BibitemOpen
  \bibfield  {author} {\bibinfo {author} {\bibfnamefont {P.}~\bibnamefont
  {Cavadini}}, \bibinfo {author} {\bibfnamefont {J.}~\bibnamefont {Krenn}},
  \bibinfo {author} {\bibfnamefont {P.}~\bibnamefont {Scharfer}},\ and\
  \bibinfo {author} {\bibfnamefont {W.}~\bibnamefont {Schabel}},\ }\bibfield
  {title} {\enquote {\bibinfo {title} {Investigation of surface deformation
  during drying of thin polymer films due to {M}arangoni convection},}\ }\href
  {https://doi.org/10.1016/j.cep.2012.11.008} {\bibfield  {journal} {\bibinfo
  {journal} {Chemical Engineering and Processing: Process Intensification}\
  }\textbf {\bibinfo {volume} {64}},\ \bibinfo {pages} {24--30} (\bibinfo
  {year} {2013})}\BibitemShut {NoStop}%
\bibitem [{\citenamefont {Ahlers}\ and\ \citenamefont {Xu}(2001)}]{Ahlers2001}%
  \BibitemOpen
  \bibfield  {author} {\bibinfo {author} {\bibfnamefont {G.}~\bibnamefont
  {Ahlers}}\ and\ \bibinfo {author} {\bibfnamefont {X.}~\bibnamefont {Xu}},\
  }\bibfield  {title} {\enquote {\bibinfo {title} {Prandtl-number dependence of
  heat transport in turbulent {R}ayleigh--{B}énard convection},}\ }\href
  {https://doi.org/10.1103/physrevlett.86.3320} {\bibfield  {journal} {\bibinfo
   {journal} {Physical Review Letters}\ }\textbf {\bibinfo {volume} {86}},\
  \bibinfo {pages} {3320--3323} (\bibinfo {year} {2001})}\BibitemShut {NoStop}%
\bibitem [{\citenamefont {Tönsmann}, \citenamefont {Scharfer},\ and\
  \citenamefont {Schabel}(2021)}]{Toensmann2021}%
  \BibitemOpen
  \bibfield  {author} {\bibinfo {author} {\bibfnamefont {M.}~\bibnamefont
  {Tönsmann}}, \bibinfo {author} {\bibfnamefont {P.}~\bibnamefont
  {Scharfer}},\ and\ \bibinfo {author} {\bibfnamefont {W.}~\bibnamefont
  {Schabel}},\ }\bibfield  {title} {\enquote {\bibinfo {title} {Transient
  three-dimensional flow field measurements by means of 3{D} µptv in drying
  poly(vinyl acetate)-methanol thin films subject to short-scale {M}arangoni
  instabilities},}\ }\href {https://doi.org/10.3390/polym13081223} {\bibfield
  {journal} {\bibinfo  {journal} {Polymers}\ }\textbf {\bibinfo {volume}
  {13}},\ \bibinfo {pages} {1223} (\bibinfo {year} {2021})}\BibitemShut
  {NoStop}%
\bibitem [{\citenamefont {Roubíček}(2021)}]{Roubicek2021}%
  \BibitemOpen
  \bibfield  {author} {\bibinfo {author} {\bibfnamefont {T.}~\bibnamefont
  {Roubíček}},\ }\bibfield  {title} {\enquote {\bibinfo {title} {From
  quasi-incompressible to semi-compressible fluids},}\ }\href
  {https://doi.org/10.3934/dcdss.2020414} {\bibfield  {journal} {\bibinfo
  {journal} {Discrete and Continuous Dynamical Systems - Series S}\ }\textbf
  {\bibinfo {volume} {14}},\ \bibinfo {pages} {4069} (\bibinfo {year}
  {2021})}\BibitemShut {NoStop}%
\bibitem [{\citenamefont {Kolegov}(2018)}]{Kolegov2018113}%
  \BibitemOpen
  \bibfield  {author} {\bibinfo {author} {\bibfnamefont {K.~S.}\ \bibnamefont
  {Kolegov}},\ }\bibfield  {title} {\enquote {\bibinfo {title} {Simulation of
  patterned glass film formation in the evaporating colloidal liquid under {IR}
  heating},}\ }\href {https://doi.org/10.1007/s12217-017-9587-0} {\bibfield
  {journal} {\bibinfo  {journal} {Microgravity Science and Technology}\
  }\textbf {\bibinfo {volume} {30}},\ \bibinfo {pages} {113--120} (\bibinfo
  {year} {2018})}\BibitemShut {NoStop}%
\bibitem [{\citenamefont {Cavadini}\ \emph {et~al.}(2015)\citenamefont
  {Cavadini}, \citenamefont {Erz}, \citenamefont {Sachsenheimer}, \citenamefont
  {Kowalczyk}, \citenamefont {Willenbacher}, \citenamefont {Scharfer},\ and\
  \citenamefont {Schabel}}]{Cavadini2015}%
  \BibitemOpen
  \bibfield  {author} {\bibinfo {author} {\bibfnamefont {P.}~\bibnamefont
  {Cavadini}}, \bibinfo {author} {\bibfnamefont {J.}~\bibnamefont {Erz}},
  \bibinfo {author} {\bibfnamefont {D.}~\bibnamefont {Sachsenheimer}}, \bibinfo
  {author} {\bibfnamefont {A.}~\bibnamefont {Kowalczyk}}, \bibinfo {author}
  {\bibfnamefont {N.}~\bibnamefont {Willenbacher}}, \bibinfo {author}
  {\bibfnamefont {P.}~\bibnamefont {Scharfer}},\ and\ \bibinfo {author}
  {\bibfnamefont {W.}~\bibnamefont {Schabel}},\ }\bibfield  {title} {\enquote
  {\bibinfo {title} {Investigation of the flow field in thin polymer films due
  to inhomogeneous drying},}\ }\href
  {https://doi.org/10.1007/s11998-015-9725-9} {\bibfield  {journal} {\bibinfo
  {journal} {Journal of Coatings Technology and Research}\ }\textbf {\bibinfo
  {volume} {12}},\ \bibinfo {pages} {921--926} (\bibinfo {year}
  {2015})}\BibitemShut {NoStop}%
\bibitem [{\citenamefont {Gerasimov}\ and\ \citenamefont
  {Yurin}(2018)}]{Gerasimov2018}%
  \BibitemOpen
  \bibfield  {author} {\bibinfo {author} {\bibfnamefont {D.~N.}\ \bibnamefont
  {Gerasimov}}\ and\ \bibinfo {author} {\bibfnamefont {E.~I.}\ \bibnamefont
  {Yurin}},\ }\href {https://doi.org/10.1007/978-3-319-96304-4} {\emph
  {\bibinfo {title} {Kinetics of evaporation}}}\ (\bibinfo  {publisher}
  {Springer International Publishing},\ \bibinfo {year} {2018})\BibitemShut
  {NoStop}%
\bibitem [{\citenamefont {Sazhin}(2006)}]{Sazhin2006}%
  \BibitemOpen
  \bibfield  {author} {\bibinfo {author} {\bibfnamefont {S.~S.}\ \bibnamefont
  {Sazhin}},\ }\bibfield  {title} {\enquote {\bibinfo {title} {Advanced models
  of fuel droplet heating and evaporation},}\ }\href
  {https://doi.org/10.1016/j.pecs.2005.11.001} {\bibfield  {journal} {\bibinfo
  {journal} {Progress in Energy and Combustion Science}\ }\textbf {\bibinfo
  {volume} {32}},\ \bibinfo {pages} {162--214} (\bibinfo {year}
  {2006})}\BibitemShut {NoStop}%
\bibitem [{\citenamefont {Tarasevich}, \citenamefont {Vodolazskaya},\ and\
  \citenamefont {Isakova}(2011)}]{Tarasevich2011}%
  \BibitemOpen
  \bibfield  {author} {\bibinfo {author} {\bibfnamefont {Y.~Y.}\ \bibnamefont
  {Tarasevich}}, \bibinfo {author} {\bibfnamefont {I.~V.}\ \bibnamefont
  {Vodolazskaya}},\ and\ \bibinfo {author} {\bibfnamefont {O.~P.}\ \bibnamefont
  {Isakova}},\ }\bibfield  {title} {\enquote {\bibinfo {title} {Desiccating
  colloidal sessile drop: dynamics of shape and concentration},}\ }\href
  {https://doi.org/10.1007/s00396-011-2418-8} {\bibfield  {journal} {\bibinfo
  {journal} {Colloid and Polymer Science}\ }\textbf {\bibinfo {volume} {289}},\
  \bibinfo {pages} {1015--1023} (\bibinfo {year} {2011})}\BibitemShut {NoStop}%
\bibitem [{\citenamefont {Patochkina}, \citenamefont {Kazarinov},\ and\
  \citenamefont {Tkachenko}(2016)}]{Patochkina2016}%
  \BibitemOpen
  \bibfield  {author} {\bibinfo {author} {\bibfnamefont {O.~L.}\ \bibnamefont
  {Patochkina}}, \bibinfo {author} {\bibfnamefont {Y.~G.}\ \bibnamefont
  {Kazarinov}},\ and\ \bibinfo {author} {\bibfnamefont {V.~I.}\ \bibnamefont
  {Tkachenko}},\ }\bibfield  {title} {\enquote {\bibinfo {title} {Physical
  model of the dependence of the nusselt number on the rayleigh number},}\
  }\href {https://doi.org/10.1134/s1063784216110177} {\bibfield  {journal}
  {\bibinfo  {journal} {Technical Physics}\ }\textbf {\bibinfo {volume} {61}},\
  \bibinfo {pages} {1626--1632} (\bibinfo {year} {2016})}\BibitemShut {NoStop}%
\bibitem [{\citenamefont {Gatapova}\ \emph {et~al.}(2014)\citenamefont
  {Gatapova}, \citenamefont {Semenov}, \citenamefont {Zaitsev},\ and\
  \citenamefont {Kabov}}]{Gatapova2014}%
  \BibitemOpen
  \bibfield  {author} {\bibinfo {author} {\bibfnamefont {E.~Y.}\ \bibnamefont
  {Gatapova}}, \bibinfo {author} {\bibfnamefont {A.~A.}\ \bibnamefont
  {Semenov}}, \bibinfo {author} {\bibfnamefont {D.~V.}\ \bibnamefont
  {Zaitsev}},\ and\ \bibinfo {author} {\bibfnamefont {O.~A.}\ \bibnamefont
  {Kabov}},\ }\bibfield  {title} {\enquote {\bibinfo {title} {Evaporation of a
  sessile water drop on a heated surface with controlled wettability},}\ }\href
  {https://doi.org/10.1016/j.colsurfa.2013.05.046} {\bibfield  {journal}
  {\bibinfo  {journal} {Colloids and Surfaces A: Physicochemical and
  Engineering Aspects}\ }\textbf {\bibinfo {volume} {441}},\ \bibinfo {pages}
  {776--785} (\bibinfo {year} {2014})}\BibitemShut {NoStop}%
\end{thebibliography}%

\end{document}